\documentclass[double]{article}
\usepackage{times}
\usepackage{algorithm}
\usepackage{algorithmic}
\usepackage{wrapfig}
\usepackage{verbatim}
\usepackage{amsmath}
\usepackage{graphicx}
\usepackage{subcaption}
\usepackage[active]{srcltx}
\usepackage{color}
\usepackage{lineno}
\usepackage{dirtytalk}
\usepackage{amsthm}
\usepackage{authblk}
\usepackage{listings}
\usepackage{tikz}
\usepackage{longtable}
\usetikzlibrary{shapes.geometric, arrows} 

\usepackage{epsfig,graphicx,amsfonts}
\usepackage{amsmath,amsthm,amssymb}
\usepackage{xcolor}

\usepackage{adjustbox}
\usepackage{multirow}
\usepackage{setspace}
\usepackage{hyperref}
\usepackage{comment}

\long\def\comment #1\commentend{}

\begin{document}

\title{\Large Friends, Foes, and First Authors: A Game Theory Model of How Power Plays Rewrite Academic Co-Authorship Networks}

\author{Amit Bengal$^{1,*}$, Teddy Lazebnik$^{2,3}$\\
\(^1\) Department of Information Sciences,  Bar Ilan University, Ramat Gan, Israel\\
\(^2\) Department of Information Science,  University of Haifa, Haifa, Israel\\
\(^3\) Department of Computing, Jonkoping University, Jonkoping, Sweden \\
\(^*\) Corresponding author: \url{amit.bengiat10@gmail.com} 
}

\date{ }

\maketitle 

\begin{abstract}
\noindent
Scientific research increasingly depends on multi-author collaboration, yet the systems used to allocate authorship credit remain vulnerable to conflict, strategic behavior, and project breakdown. Although prior work has shown that authors may rationally issue ultimatums over authorship order within a single manuscript, much less is known about how such behavior unfolds over repeated collaborations embedded in evolving academic networks. In this study, we develop a repeated, networked game-theoretic model of co-authorship in which researchers form collaborations over time, accumulate reputation through an evolving friendship network, and, in a subset of cases, learn strategic behavior through deep reinforcement learning. Using large-scale agent-based simulations, we compare myopic and forward-looking authors across mixed populations. We find that strategic agents do not raise fewer ultimatums than greedy agents, but instead learn to avoid insisting after rejection, thereby eliminating destructive manuscript termination. As strategic prevalence increases, paper destruction falls from 0.120 to 0.000 per paper, completion rates rise from 0.853 to 0.970, and average completed papers per agent increase from 15.2 to 16.9. Strategic agents also obtain a substantial utility advantage, reaching 30.8\% when rare, while overall inequality remains stable. These results suggest that reputational feedback and long-term incentives can make academic collaboration more resilient, offering a computational testbed for designing fairer and more productive authorship policies. \\ \\

\noindent
\textbf{Keywords}: co-authorship networks; authorship disputes; ultimatum game; academic collaboration; reputation dynamics; deep reinforcement learning; agent-based modeling; scientific productivity.
\end{abstract}

\maketitle \thispagestyle{empty}
\pagestyle{myheadings} \markboth{Draft:  \today}{Draft:  \today}
\setcounter{page}{1}

\section{Introduction}
\label{sec:introduction}
Scientific collaboration has become a dominant mode of knowledge production across almost all academic fields \cite{dehdarirad2017research,de2021measure,spreaded}. The proportion of multi-authored manuscripts has steadily increased over recent decades, and team-authored work is now the norm rather than the exception \cite{wuchty2007increasing,amjad2017standing}. This trend is driven by multiple forces, including the growing complexity of research questions, the need for complementary expertise, and institutional incentives that reward productivity and visibility \cite{acedo2006co,zhang2018understanding}. This trend is further amplified by the rise of interdisciplinary research, which has grown exponentially as academic boundaries become increasingly blurred and funding agencies prioritize cross-disciplinary collaborations \cite{von2017academe}.

However, alongside the celebrated benefits of co-authorship \cite{katz1997you}, a growing body of work has documented its \say{dark side}, including exploitation \cite{lawrence2002rank}, bullying \cite{avila2014bullying}, unethical authorship practices \cite{je2021sidelined, street2010credit}, and disputes over credit allocation \cite{primack2014editorial}. Alerting enough, \cite{borry2006author} shows that there is a linear relationship between the number of authors involved in a project and the magnitude of these challenges, indicating that the growing number of multi-author studies is undermining the effectiveness of the long-standing scientific authorship system. Recently, \cite{savchenko2024authorship} report results from an international cross-discipline survey of 752 academics from 41 fields in 93 countries, showing that conflicts over authorship credit arise already at the Master’s and PhD stages, affect around a quarter of advisees, and are experienced with peers by roughly half of respondents, suggesting that authorship disputes are a widespread, systemic feature of contemporary science rather than a problem of a few \say{bad apples}. 

Due to the widespread nature of this phenomenon, it is important to understand its roots in order to allow policymakers, as well as, scholars in the community to better navigate and alter these dynamics \cite{newman2001structure}. One approach to such a task is using mathematical modeling, in general, and game-theory based models, in particular \cite{jackson1996strategic}. For instance, \cite{carayol2009knowledge} developed a strategic model examining how researchers choose collaborators based on trade-offs between communication costs and knowledge gains. In a similar manner, \cite{bobtcheff2017researcher} modeled the strategic decisions researchers face when allocating effort between safe and innovative research projects. \cite{bikard2015exploring} used game theory to analyze the tensions between collaboration benefits and individual credit allocation in team science.

Specifically, \cite{lazebnik2023academic} formalized one important facet of these tensions by modeling academic co-authorship as an ultimatum game over the authorship order of a single manuscript. In their model, rational authors repeatedly contribute to a shared project and may, at any point prior to submission, issue an ultimatum demanding an improved position on the authors' list. If all co-authors accept, the authorship order is updated; if at least one co-author rejects, the ultimatum-issuing author may either withdraw (and incur a penalty) or insist, thereby terminating the project and destroying everyone's gains. Using an agent-based implementation of this game, the authors showed that the probability that at least one author finds it rational to raise such an ultimatum is surprisingly high, even under seemingly benign conditions such as equal contributions and utilities. The likelihood of ultimatum issuance increases with the number of authors, the heterogeneity in contributions, and certain configurations of project duration and stage. While this one-shot, single-project model already yields disturbing insights about the structural fragility of current co-authorship practices, it abstracts away several central features of real academic life. First, collaborations are rarely isolated events as researchers co-author multiple projects over time, often with overlapping subsets of colleagues. Second, collaborations are embedded in social and professional networks in which reputation, trust, and prior experience strongly shape future opportunities. A destructive ultimatum that might be locally advantageous in a single project can propagate through the collaboration network, influencing who is willing to work with whom in the future. Third, the decision to collaborate is itself strategic as researchers can prefer familiar, trusted partners or invest in exploring new connections, balancing short-term risk with long-term diversification \cite{ross2010challenges}.

To this end, in this study, we extend the original co-authorship ultimatum model into a repeated, networked game of academic collaboration. This dynamic framework allows us to move beyond the question of whether issuing an ultimatum is rational in a single project, and instead ask when such behavior is sustainable in the long run in an interconnected network of individuals. Concretely, we study how network structure, exploration tendencies, and institutional parameters governing rewards and penalties shape the emergent prevalence of ultimatums, the distribution of cumulative utilities, and the fragmentation of the collaboration network over time. Our main research questions are:
\begin{enumerate}
    \item \textbf{RQ1:} Under what conditions does the possibility of future collaborations and reputational spillovers discourage locally optimal ultimatums.
    \item \textbf{RQ2:} When forward-looking agents coexist with myopic ones, who benefits, and at whose expense.
\end{enumerate}
In order to address these questions, we implemented the proposed model as an agent-based simulation using real-world settings from the literature. We then conduct a series of \textit{in silico} experiments that compare the original single-project model to its networked extension, revealing how local incentives interact with global structure to produce qualitatively different collaboration patterns at the community level.

The rest of this manuscript is organized as follows. Section \ref{sec:model} formally introduced the proposed model. Section \ref{sec:experiments} outlines the experiments conducted using the model and their results. Section \ref{sec:discussion} discusses the applicative implicative of these results, mentions the study's limitations, and suggests possible future work.

\section{Model Definition}
\label{sec:model}
The proposed model is constructed via three interconnected components. Namely, the connection network that defines the topology of the model alongside its construction, its dynamics over time following the \say{ultimatum raising game}, the author's utility, and the author's strategic decision-making mechanism. Below, we formally define each of these components and their interactions.

\subsection{Network topology and construction}
Let us consider a dynamic co-authorship game played by a fixed set of researchers $\mathcal{P} = \{p_1,\dots,p_n\}$, where $n := |\mathcal{P}| < \infty$ is the population size. Time evolves in discrete steps $t = 0,1,\dots,T$, where $T < \infty$ is the time horizon of the game. At any time $t$, some researchers are engaged in one or more ongoing papers, while others are idle and available to form new collaborations. The state of the system at time $t$ is represented by two intertwined networks: a persistent friendship network that encodes long-term social and professional ties, and a dynamic collaboration network describing the set of currently active papers. 

The friendship network at time $t$ is a weighted, undirected graph $G_{\text{friendship}}^{(t)} = (\mathcal{P}, E_{\text{friendship}}^{(t)}, W^{(t)})$. The vertex set, \(V\), coincides with the set of researchers $\mathcal{P}$. An unordered pair $\{p_i,p_j\}$ belongs to $E_{\text{friendship}}^{(t)}$ if and only if the two researchers maintain a direct friendship or professional tie at time $t$. Each existing edge $\{p_i,p_j\}$ carries a weight $w_{ij}^{(t)} \in (0,1]$ that represents the current strength of their relationship, such that larger values correspond to stronger ties. 

At initialization, $t = 0$, the friendship network is generated as follows. Each researcher $p_i$ draws a desired degree $\deg(p_i)$ from a Poisson distribution with parameter $\lambda > 0$, which controls the expected number of friendships. The researcher then selects $\min(\deg(p_i), n-1)$ distinct partners from $\mathcal{P} \setminus \{p_i\}$ uniformly at random. Mutual selections result in a single undirected edge, so that $\{p_i,p_j\} \in E_{\text{friendship}}^{(0)}$ if at least one of $p_i$ or $p_j$ has selected the other. The weight of each initial friendship edge is sampled from a truncated normal distribution: for every $\{p_i,p_j\} \in E_{\text{friendship}}^{(0)}$ we draw $\tilde{w}_{ij} \sim \mathcal{N}(\mu_w,\sigma_w^2)$ and set $w_{ij}^{(0)} = \min\{1,\max\{0,\tilde{w}_{ij}\}\}$ so that $w_{ij}^{(0)} \in (0,1]$. 

In a complementary manner, the collaboration network at time $t$ is a hypergraph $G_{\text{collab}}^{(t)} = (\mathcal{P}, \mathcal{C}^{(t)})$, where $\mathcal{C}^{(t)} = \{\mathcal{K}_1^{(t)}, \dots, \mathcal{K}_{M_t}^{(t)}\}$ is the set of collaboration cliques corresponding to papers in progress. Each clique $\mathcal{K}_m^{(t)} \subseteq \mathcal{P}$ is the set of co-authors currently working together on paper $m$ at time $t$, and $M_t$ denotes the number of active papers. 

Clique formation is driven by the friendship network and by individual exploration tendencies. Each researcher $p_i$ is endowed with an exploration parameter $e_i \in [0,1]$ that governs the probability of seeking new, potentially distant collaborators versus reinforcing existing local ties. The effective strength of the relationship between two researchers $p_i$ and $p_j$ in the friendship network at time $t$ is quantified by a path-based measure. Let $\mathcal{P}_{i \to j}^{\mathrm{sp}}$ be any shortest path between $p_i$ and $p_j$ in $G_{\text{friendship}}^{(t)}$ in terms of number of hops, and denote by $w_{uv}^{(t)}$ the weight of edge $\{u,v\}$ along this path. We define
\[
    d_G^{(t)}(p_i,p_j)
    \;=\;
    \prod_{\{u,v\} \in \mathcal{P}_{i \to j}^{\mathrm{sp}}} w_{uv}^{(t)},
\]
so that $d_G^{(t)}(p_i,p_j) \in (0,1]$ captures the multiplicative strength of the best (strongest) chain of friendships connecting the two researchers. Researchers that are not connected in $G_{\text{friendship}}^{(t)}$ have $d_G^{(t)}(p_i,p_j) = 0$ by convention and cannot be selected as collaborators in the current step. 

When a new paper $m$ is about to be formed, a seed author $p_i$ is selected among the idle researchers according to some exogenous rule (e.g., uniformly at random or proportional to remaining capacity). The target clique size $|\mathcal{K}_m^{(t)}|$ is then sampled in two stages. First, a Poisson random variable $X \sim \mathrm{Pois}(\lambda_K)$ is drawn. Second, the realized size is projected to the admissible range by setting
\[
    |\mathcal{K}_m^{(t)}| = \min\{\max(X,2), K\},
\]
where $K \ge 2$ is the maximal allowable clique size. This ensures that each paper has at least two authors and at most $K$ co-authors. Starting from the seed $p_i$, additional members are recruited iteratively until the clique reaches the target size or available candidates are exhausted. At each recruitment step, the current focal author $p_r$ chooses the next collaborator according to an exploration--exploitation rule. With probability $e_r$, the author explores by sampling a candidate $p_j$ from all reachable researchers (excluding current clique members and saturated agents) with probability proportional to $d_G^{(t)}(p_r,p_j)$, i.e.,
\[
    \Pr(p_j \mid \text{exploration}, p_r)
    \;=\;
    \frac{d_G^{(t)}(p_r,p_j)}{\sum_{k \neq r} d_G^{(t)}(p_r,p_k)}.
\]
With complementary probability $1 - e_r$, the author exploits local ties by restricting attention to immediate neighbors in the friendship network and sampling uniformly or in proportion to the edge weights $w_{rk}^{(t)}$ over neighbors $p_k$ with remaining capacity. If at any stage the focal author has no feasible candidates (for instance, because all neighbors are already overloaded or part of the current clique), the algorithm backtracks to the previous clique member and resamples a different collaborator. If the sampled clique size exceeds the number of researchers with available capacity, the clique is truncated and the paper is formed from the maximal feasible subset. The resulting set of authors defines a new collaboration clique $\mathcal{K}_m^{(t)}$, which is added to $\mathcal{C}^{(t)}$.

\begin{figure}
    \centering
    \includegraphics[width=0.99\linewidth]{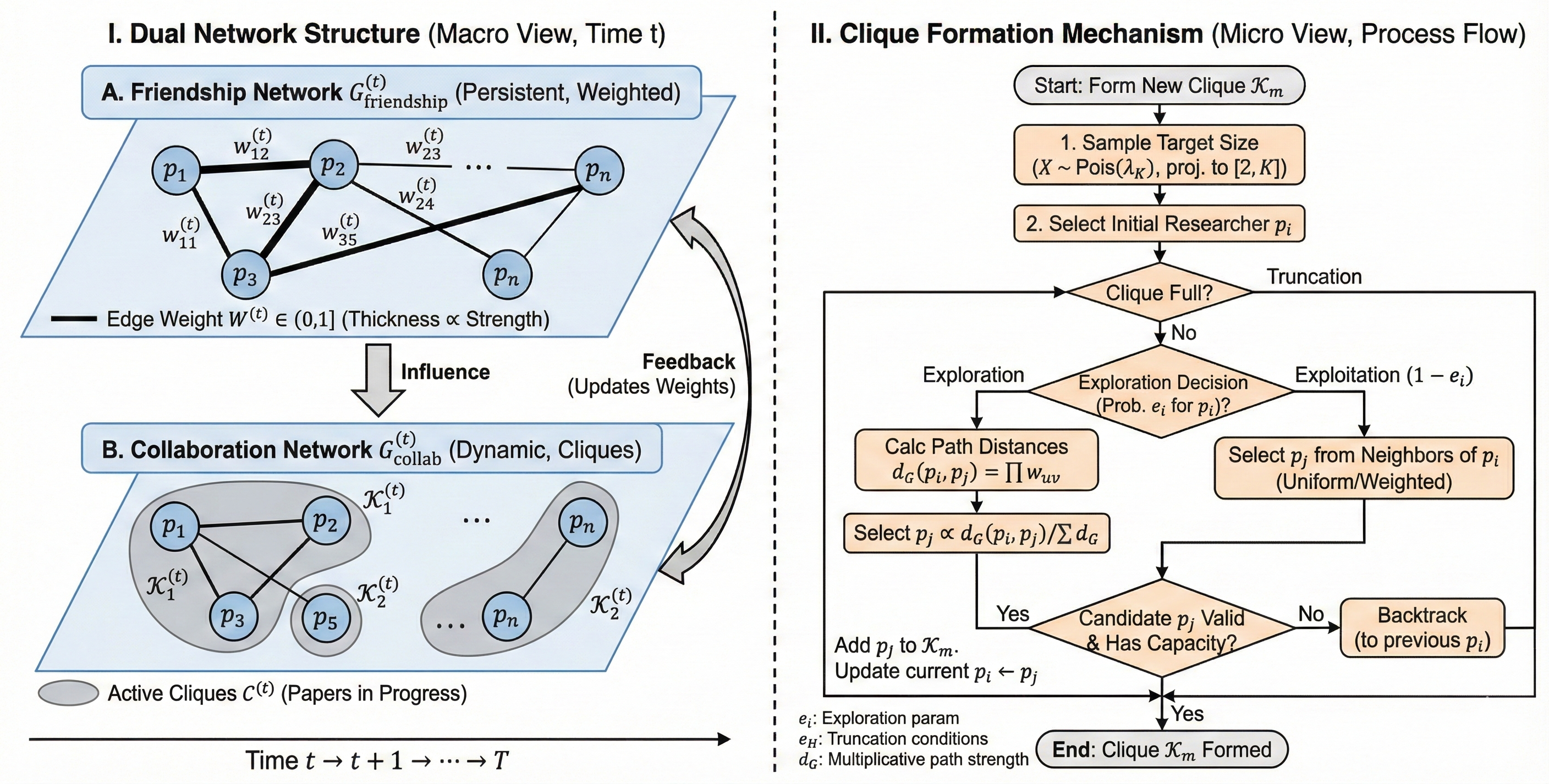}
    \caption{A schematic view of the proposed model.}
    \label{fig:scheme}
\end{figure}

Fig. \ref{fig:scheme} presents a schematic view of (i) the network and (ii) the process that constructs graph cliques from the friendship graph.

\subsection{Ultimatum raising game}
Within each clique, the collaboration on paper $m$ unfolds over a sequence of discrete steps $\tau = 1,\dots,T_m$ aligned with the global clock, where the duration $T_m \in \mathbb{N}$ is a random variable sampled from a discrete uniform distribution $\mathcal{U}(t_1,t_2)$ such that \(t_2 > t_1\). For each author $p_i$ participating in paper $m$, we define a project-specific utility $u_{i,m}^0 \in \mathbb{R}^+$ for successful completion of the manuscript and an authorship-position utility function $u_{i,m}^1 : \{1,\dots,|\mathcal{K}_m^{(t)}|\} \to [0,1]$. Following the original model \cite{lazebnik2023academic}, the authorship-position utility is assumed to be strictly decreasing and concave in the position index $j$, so that
\begin{equation}
    \frac{\partial u_{i,m}^1(j)}{\partial j} < 0
    \quad\text{and}\quad
    \frac{\partial^2 u_{i,m}^1(j)}{\partial j^2} < 0.
\end{equation}
The overall utility obtained by author $p_i$ from being in position $j$ on the final authors' list of paper $m$ is therefore $u_{i,m}^0 \cdot u_{i,m}^1(j)$. During week $\tau$ of paper $m$, each author $p_i \in \mathcal{K}_m^{(t)}$ contributes a fraction $\beta_{i,m}^{(\tau)}$ of the total effort. The contributions are normalized so that
\begin{equation}
    \sum_{\tau = 1}^{T_m} \sum_{p_i \in \mathcal{K}_m^{(t)}} \beta_{i,m}^{(\tau)} = 1,
\end{equation}
and the cumulative contribution of $p_i$ to paper $m$ at the time of completion or termination is denoted by
\begin{equation}
    \text{contrib}_{i,m} = \sum_{\tau=1}^{\tau_m^\star} \beta_{i,m}^{(\tau)},
\end{equation}
where $\tau_m^\star$ is the stopping time of the project.

At any week $\tau < T_m$, an author $p_i$ may attempt to improve their authorship position by issuing an ultimatum. Suppose that at week $\tau$ the current authors' list for paper $m$ is an ordered list of the members of $\mathcal{K}_m^{(t)}$, and that $p_i$ currently occupies position $j$. The author may propose to move to a strictly better position $j' < j$ by threatening to block the submission of the manuscript if their demand is not met. The other authors then simultaneously and independently decide whether to accept or reject the ultimatum. If all co-authors accept, the ultimatum is successful and the authorship order is updated: $p_i$ moves to position $j'$, and authors currently at positions $j',\dots,j-1$ are shifted one position downward. If at least one co-author rejects, the ultimatum fails in the short run and $p_i$ must decide whether to withdraw or maintain the ultimatum. Withdrawing preserves the project but imposes a penalty on $p_i$ (for example, through a reduction in their effective contribution or utility), while maintaining the ultimatum terminates the paper $m$ and yields negative utilities for all participants proportional to their invested contributions. By default, within-paper decision logic follows the structure of the ultimatum game introduced in \cite{lazebnik2023academic}, which is greedy and does not explicitly aims to optimize the cumulative utility of the authors. In the next sub-section, we propose a decision-making data-driven model associated with a \say{strategic} author.

The outcome of each paper induces an update of the friendship network, thereby linking local collaboration events to the global co-authorship structure. When paper $m$ is successfully completed (either without ultimatum or with an accepted ultimatum), the clique $\mathcal{K}_m^{(t)}$ is removed from the collaboration network, i.e., $\mathcal{K}_m^{(t)}$ is deleted from $\mathcal{C}^{(t)}$. For every unordered pair $\{p_i,p_j\} \subseteq \mathcal{K}_m^{(t)}$, $i \neq j$, we then strengthen or create a friendship tie. If $\{p_i,p_j\} \in E_{\text{friendship}}^{(t)}$, its weight is increased according to
\[
    w_{ij}^{(t+1)} = \min\bigl(1,\, w_{ij}^{(t)} + \delta_{\text{success}}\bigr),
\]
where $\delta_{\text{success}} > 0$ controls the incremental reinforcement. If no edge previously existed between $p_i$ and $p_j$, we add the edge to the friendship set and initialize its weight by drawing $w_{ij}^{(t+1)} \sim \mathcal{U}(0,1]$. All other edges and weights are kept unchanged.

If an ultimatum is issued, rejected by at least one co-author, and subsequently withdrawn by $p_i$, the paper continues but the relationships within the clique deteriorate. In this case, the collaboration network remains unchanged, while the friendship weights between $p_i$ and the other members of $\mathcal{K}_m^{(t)}$ are updated according to
\[
    w_{ij}^{(t+1)} =
    \begin{cases}
        \max\bigl(0,\, w_{ij}^{(t)} - \delta_{\text{withdraw}}\bigr) & \text{if } p_j \in \mathcal{K}_m^{(t)} \setminus \{p_i\}, \\
        w_{ij}^{(t)} & \text{otherwise},
    \end{cases}
\]
where $\delta_{\text{withdraw}} > 0$ is a small penalty parameter. This captures the idea that even a failed attempt to renegotiate credit can leave scars on local relationships without necessarily destroying the project.

Finally, if $p_i$ maintains a rejected ultimatum, paper $m$ is terminated and removed from $\mathcal{C}^{(t)}$. Direct relationships between $p_i$ and their co-authors in $\mathcal{K}_m^{(t)}$ are set to zero, i.e.,
\[
    w_{ij}^{(t+1)} = 0
    \quad\text{for all } p_j \in \mathcal{K}_m^{(t)} \setminus \{p_i\},
\]
effectively severing these ties. To model reputational spillovers beyond the immediate clique, we further propagate penalties along the friendship network. For each researcher $p_j$ outside the clique who is connected, directly or indirectly, to at least one member of $\mathcal{K}_m^{(t)}$, we define
\[
    d_G^{(t)}(p_j,\mathcal{K}_m^{(t)}) \;=\;
    \min_{p_\ell \in \mathcal{K}_m^{(t)}} d_G^{(t)}(p_j,p_\ell),
\]
that is, the strongest path-based connection from $p_j$ to the clique. Let
\[
    \phi_{ij}^{(t)} \;=\;
    \Bigl[\gamma + \theta \cdot 
    \max_{p_\ell \in \mathcal{K}_m^{(t)} \setminus \{p_i\}}
    \bigl(\text{contrib}_{\ell,m}\bigr)\Bigr]
    \cdot 
    \exp\bigl(-\alpha \cdot d_{G}^{(t)}(p_j,\mathcal{K}_m^{(t)})\bigr),
\]
where $\gamma > 0$ is a base reputation penalty for destructive behavior, $\theta > 0$ scales the penalty with the maximum effort wasted by other clique members, and $\alpha > 0$ controls the rate at which reputational damage decays with network distance. Fix a small threshold $\varepsilon > 0$. The friendship weight $w_{ij}^{(t)}$ is then updated recursively as
\[
    w_{ij}^{(t+1)} =
    \begin{cases}
        (1 - \phi_{ij}^{(t)}) \, w_{ij}^{(t)}, & \text{if } w_{ij}^{(t)} > \varepsilon, \\[6pt]
        0, & \text{if } w_{ij}^{(t)} \le \varepsilon,
    \end{cases}
\]
with $w_{ij}^{(t)} \in [0,1]$ for all $t$. Once $w_{ij}^{(t)}$ falls below $\varepsilon$, the edge is effectively removed from the friendship network, and subsequent updates leave it at zero. This mechanism enforces that destructive ultimata can have long-range consequences that gradually erode collaboration opportunities for the initiating author.

\subsection{Author utility}
Over the entire horizon $t = 0,\dots,T$, each researcher may participate in multiple papers. Let $\mathcal{M}_i$ denote the set of indices of papers that researcher $p_i$ successfully completes:
\[
    \mathcal{M}_i \;=\;
    \{ m : p_i \in \mathcal{K}_m^{(t)} \text{ for some } t,\ \text{and paper } m \text{ was completed} \}.
\]
If $j_{i,m}$ is the final authorship position of $p_i$ on paper $m$, and $t_m$ is the macro-time at which paper $m$ is completed, then the cumulative discounted utility of researcher $p_i$ over the entire game is
\[
    U_i^{\text{total}} =
    \sum_{m \in \mathcal{M}_i}
    \frac{u_{i,m}^0 \cdot u_{i,m}^1(j_{i,m})}{(1 + \rho)^{t_m}},
\]
where $\rho \in [0,1]$ is a discount factor that captures time preferences or the diminishing marginal value of future publications. 

\subsection{Strategic authors}
In the original \say{ultimatum raising game}, each author $p_i$ decides in a greedy manner whether to issue an ultimatum and either to accept it or not in a given paper, optimizing only the utility from that single manuscript, ignoring the long-term outcomes following the specific paper. However, in reality, it is safe to assume authors are more strategic in their behavior. Namey, strategic authors' goal is to learn a policy that maximizes the long–run cumulative utility $U^{\text{total}}_i$ by internalizing the impact of ultimata on future collaboration opportunities and reputation in the friendship network. 

To that end, we formulate the ultimatum decision of each agent as a Deep Reinforcement Learning (DRL) task \cite{arulkumaran2017deep,lazebnik2023data}. Concretely, we model the simulation as a Markov decision process \cite{sutton1998reinforcement} in which, at each decision time $t$, an agent $p_i$ observes a continuous, high–dimensional state vector $s_t^i$ summarizing the current paper, their own publication history, and their local position in the friendship network; based on this state, the agent selects a discrete action $a_t^i$ (issue an ultimatum or not), receives a scalar reward $r_t^i$ reflecting realized publication utility and reputational consequences, and transitions to a new state $s_{t+1}^i$ as the environment (i.e., the agent-based simulation) evolves. Technically, a deep neural network \cite{kazak2019verifying} with parameters $\theta$ is trained to approximate either the stochastic policy $\pi_\theta(a\mid s)$ which is a function indicating the learned probability to perform an action \(a\) given the author's state \(s\), using temporal–difference learning \cite{sutton1988learning} on trajectory data generated by running the simulator. The key advantage of DRL for this task is its ability to handle the non–linear, path–dependent effects of ultimata on future collaboration opportunities \cite{arulkumaran2017deep,shuchami2025spatio}. Notably, instead of implementing a hand–crafted rule or solving a tractable special case, the agent learns directly from many simulated careers, which patterns of ultimatum behavior maximize the expected discounted sum of future utilities \cite{silver2017mastering,mazyavkina2021reinforcement}. Technically, we assume centralized training with decentralized execution where a shared DRL policy is trained from the experience of all agents, but at run time, each agent acts only on its own local state. In the Appendix, we formally introduce the model's state space, action space, architecture, training procedure, and inference procedure. 

\section{Experiments}
\label{sec:experiments}

\subsection{Setup}
We implemented the proposed model using an agent-based simulation approach \cite{wilensky2015introduction}, implemented in the Python programming language \cite{van2007python}. In order to allow comparison with the original model \cite{lazebnik2023academic}, we use similar distributions for project-level utilities and contributions. For each paper $m$ and author $p_i \in \mathcal{K}_m^{(t)}$, we draw $u_{i,m}^0$ from a bounded interval (e.g., $u_{i,m}^0 \sim \mathcal{U}[u_{\min},u_{\max}]$), and we instantiate $u_{i,m}^1(j)$ using a decreasing, concave function of the form
\[
    u_{i,m}^1(j) =
    \frac{1 - \eta_{i,m}}{j + \xi_{i,m}},
\]
where $\eta_{i,m}$ and $\xi_{i,m}$ are small random perturbations that introduce heterogeneity across authors and projects while preserving the qualitative shape assumed in Section~\ref{sec:model}. Project durations $T_m$ are drawn from a discrete uniform distribution on $[t_1,t_2]$, and the global discount factor $\rho$ is fixed. The parameters $\delta_{\text{success}}$, $\delta_{\text{withdraw}}$, $\gamma$, $\theta$, $\alpha$, and $\varepsilon$ control the strength and reach of reputational feedback and are systematically varied. Table \ref{tab:parameters} provides a summary of the model's parameters with their description and default values.

\begin{table}[ht!]
\centering
\caption{Summary of the model's parameters with their description and default values.}
\label{tab:parameters}
\begin{tabular}{llcc}
\hline \hline
\textbf{Category} & \textbf{Parameter} & \textbf{Symbol} & \textbf{Default Value} \\
\hline \hline
\multirow{3}{*}{Population} 
    & Number of agents & $n$ &  10,000 \\
    & Simulation steps & $T$ & 1,565 \\
    & Papers per round & - &  0-99,999 \\
\hline
\multirow{4}{*}{Network} 
    & Friendship degree parameter & $\lambda_{\text{friendship}}$ & 3.0 \\
    & Exploration mean & $\bar{e}$ & 0.05 \\
    & Exploration std & $\sigma_e$ & 0.02 \\
    & Rate per agent & - & 0.001 \\
\hline
\multirow{3}{*}{Collaboration} 
    & Clique size parameter & $\lambda_K$ & 3.0 \\
    & Max clique size & $K$ &  8  \\
    & Paper duration & $[t_1, t_2]$ & [8, 88] weeks \\
\hline
\multirow{5}{*}{Paper Utilities} 
    & Base utility range & $[u_{\min}, u_{\max}]$ & [10, 100] \\
    & Position utility parameter & $\eta$ & 0.5-0.8 \\
    & Position utility parameter & $\xi$ & 0.1 \\
    & Utility normalization & - & $u_0$/100 (in DRL) \\
    & Economic discount & $\rho$ & 0.05 \\
\hline
\multirow{5}{*}{Reputation} 
    & Success reinforcement & $\delta_{\text{success}}$ & 0.1 \\
    & Withdrawal penalty & $\delta_{\text{withdraw}}$ & 0.05 \\
    & Base reputation penalty & $\gamma$ & 0.2 \\
    & Contribution-scaled penalty & $\theta$ & 0.5 \\
    & Distance decay rate & $\alpha$ & 1.0 \\
\hline
\multirow{8}{*}{DRL Training} 
    & Number of strategic agents & $n_{\text{strategic}}$ & $n_{\text{strategic}}$ \\
    & Learning rate & - & $10^{-4}$ \\
    & DRL discount factor & $\gamma_{\text{RL}}$ & 0.99 \\
    & Initial epsilon & $\varepsilon_0$ & 1.0 \\
    & Final epsilon & $\varepsilon_f$ & 0.01 \\
    & Epsilon decay & - & 0.9825 \\
    & Replay buffer size & - & 10,000-100,000 \\
    & Batch size & - & 32 \\
\hline
\multirow{4}{*}{Neural Network} 
    & Paper features dim & - & 14 \\
    & Agent features dim & - & 8 \\
    & Network features dim & - & 5 \\
    & Hidden layer dim & - & 128 \\
\hline
\multirow{3}{*}{Training Schedule} 
    & Number of episodes & - & 500 \\
    & Conversion rate & $k$ & Dynamic* \\
    & Target update frequency & - & 100-1,000 \\
\hline \hline
\end{tabular}
\vspace{2mm}
\begin{flushleft}
Conversion rate computed dynamically: $k = \lceil(n - n_{\text{strategic}}) / (0.8 \times \text{episodes} / 10)\rceil$
\end{flushleft}
\end{table}

\subsection{Analysis design}
\label{sec:analysis_design}
We designed a series of analyses that are all derived from a single core experiment, namely, a mixed-population simulation in which the proportion of strategic (DRL-trained) agents is systematically varied from 0\% to 100\% in increments of 10 percentage points, while the remaining agents follow the greedy baseline policy introduced in \cite{lazebnik2023academic}. For each of the eleven resulting configurations, a full simulation of $T = 1565$ steps (corresponding to approximately 30~years of weekly decisions) is executed with $n = 10000$ agents. The analyses described below are organized so as to progressively zoom in from system-level outcomes to individual behavioral mechanisms, together providing a multi-level account of how the presence of forward-looking agents reshapes the dynamics of academic collaboration.

At the broadest level, we track a comprehensive set of core outcome indicators for each population composition. Concretely, we record the mean number of ultimatums raised per paper and their breakdown into accepted, withdrawn, and terminated outcomes; the paper completion and destruction rates; the average number of completed papers per agent; and the mean cumulative utility $\mu_U$ together with its standard deviation $\sigma_U$, computed separately for strategic agents, greedy agents, and the combined population. From these quantities, we derive the absolute and relative utility advantage of strategic agents over greedy agents, as well as the Gini coefficient of cumulative utilities within each subpopulation and overall. In order to go beyond summary statistics and capture the full shape of the utility distributions, we additionally construct a three-dimensional visualization (Fig.~\ref{fig:utility_distribution_mixed}) in which each population composition is represented by a pair of smoothed probability density curves (one for strategic agents and one for greedy agents), layered along a third axis that encodes the strategic-agent percentage. This representation makes it possible to observe not only shifts in the central tendency but also changes in the spread, skewness, and degree of overlap between the two distributions as the fraction of strategic agents grows. Taken together, these aggregate indicators and their visualization directly address \textbf{RQ2}, as they quantify who benefits from the coexistence of forward-looking and myopic agents and at whose expense, while also providing the system-level indicators required for assessing \textbf{RQ1}.

Having established this macro-level picture, we proceed to disaggregate ultimatum behavior according to the type of the initiating and responding agents, with the aim of uncovering the behavioral mechanisms that underlie the aggregate outcomes. For each population composition, we compute the per-agent initiation rate (defined as the number of ultimatums raised divided by the number of papers in which the agent participated) separately for strategic and greedy agents; the conditional outcome distribution (acceptance, withdrawal, termination) given the type of the initiator; the acceptance rate when a strategic versus a greedy agent acts as a responder to another author's ultimatum; and two escalation-control metrics that capture the \say{exit} behavior following a failed demand. The first of these is the \emph{restraint rate}, defined as the fraction of failed ultimatums (i.e., those not accepted) that ended in voluntary withdrawal rather than paper destruction. The second is the \emph{destruction rate}, defined as the fraction of all ultimatums raised by a given agent type that resulted in the termination of the project. These fine-grained metrics reveal how the DRL-trained policy modulates both the decision to raise an ultimatum and, crucially, the decision to insist or back down after rejection, thereby illuminating the specific behavioral channel through which strategic agents secure higher long-run utility. This analysis is pertinent to both \textbf{RQ1}, as it shows when reputational spillovers discourage locally optimal ultimatums, and \textbf{RQ2}, as it reveals the asymmetric costs borne by greedy agents when confronted with forward-looking opponents.

We next shift attention from the question of \emph{who} raises ultimatums to the question of \emph{when} they are raised within a paper's lifecycle. For each ultimatum event recorded across all mixed-population configurations, we store the internal week $\tau$ at which the demand was issued together with the paper's total scheduled duration $T_m$. From these data, we construct smoothed density estimates of ultimatum timing for strategic and greedy initiators separately, and compute outcome probabilities (acceptance, withdrawal, termination) as a function of binned week intervals. This temporal analysis allows us to determine whether strategic agents learn to time their demands differently from greedy agents and whether ultimatums raised early in the life of a project differ systematically from those raised late in terms of their likelihood of success. The results bear directly on \textbf{RQ1}, as they reveal how the temporal structure and the sunk-cost profile of a collaborative project interact with the incentive to renegotiate credit.

In a complementary manner, we examine whether the introduction of strategic agents alters the inequality structure of academic output at the population level. Following the classical framework of Lotka \cite{lotka1926frequency}, we compute the complementary cumulative distribution function (CCDF) of the number of completed papers per agent under each population composition, with separate curves for the strategic and greedy subpopulations. We fit both power-law and log-normal models to the tail of each distribution and report the estimated exponents, goodness-of-fit statistics ($R^2$), and the share of total production attributable to the top 10\% most productive authors. This analysis connects the proposed simulation to the well-documented skewness of scientific productivity \cite{newman2004coauthorship} and allows us to assess whether strategic behavior exacerbates or mitigates the concentration of output, which is a question with direct implications for \textbf{RQ2}.

Finally, in order to establish the structural baseline against which the effects of the DRL policy should be measured, we provide a longitudinal characterization of the simulation under the pure greedy regime (0\% strategic agents). At each time step, we record the number of active, completed, and terminated papers; the mean agent utility and its standard deviation; and a set of network-level diagnostics that include the global clustering coefficient $C(t)$, the average shortest path length $L(t)$ in the giant component, the network density, and the number of connected components. We compare the observed clustering and path-length trajectories against Erd\H{o}s--R\'{e}nyi random-graph baselines ($C_{\mathrm{ER}} \approx \langle k \rangle / N$ and $L_{\mathrm{ER}} \approx \ln N / \ln \langle k \rangle$) in order to assess whether the evolving friendship network exhibits small-world properties \cite{watts1998collective}. This baseline characterization is essential for the interpretation of the mixed-population results, as it reveals the network topology and production dynamics that emerge even in the absence of strategic reasoning. The results of all five analyses are presented in the following subsection.

\section{Results}
\label{sec:experiments_results}

\begin{figure}[H]
    \centering
    \begin{subfigure}[t]{\textwidth}
        \centering
        \includegraphics[width=\textwidth]{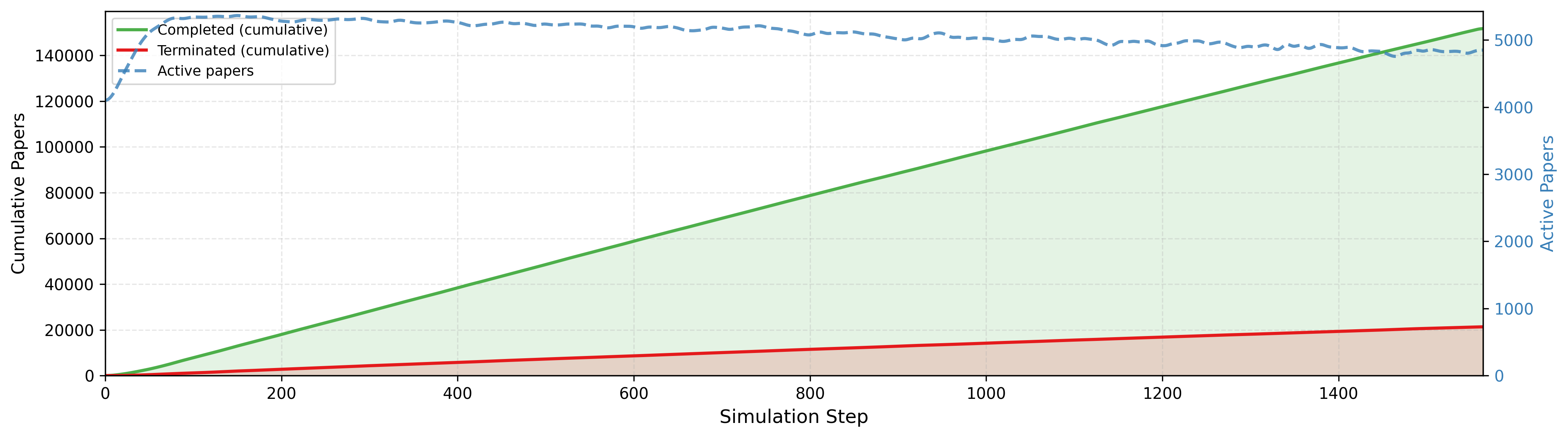}
        \caption{Cumulative completed (green) and terminated (red) papers together 
        with the number of active papers (blue, right axis).}
        \label{fig:dynamics_baseline_a}
    \end{subfigure}
    \\[1ex]
    \begin{subfigure}[t]{\textwidth}
        \centering
        \includegraphics[width=\textwidth]{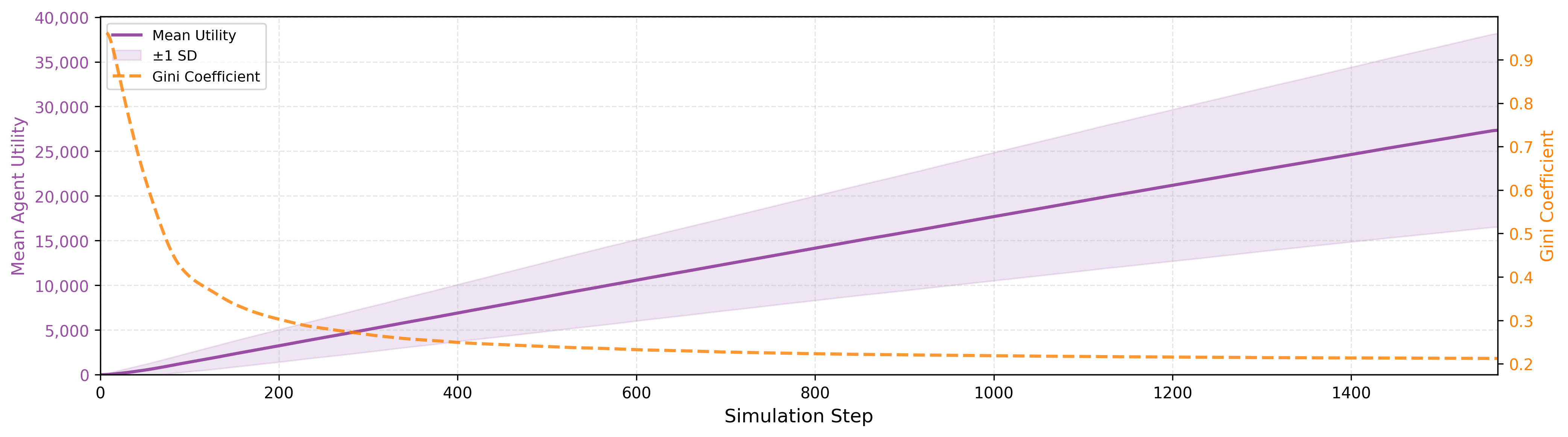}
        \caption{Mean agent utility over time (purple) with $\pm 1$ SD band, and 
        the Gini coefficient of cumulative utilities (orange, right axis).}
        \label{fig:dynamics_baseline_b}
    \end{subfigure}
    \\[1ex]
    \begin{subfigure}[t]{\textwidth}
        \centering
        \includegraphics[width=\textwidth]{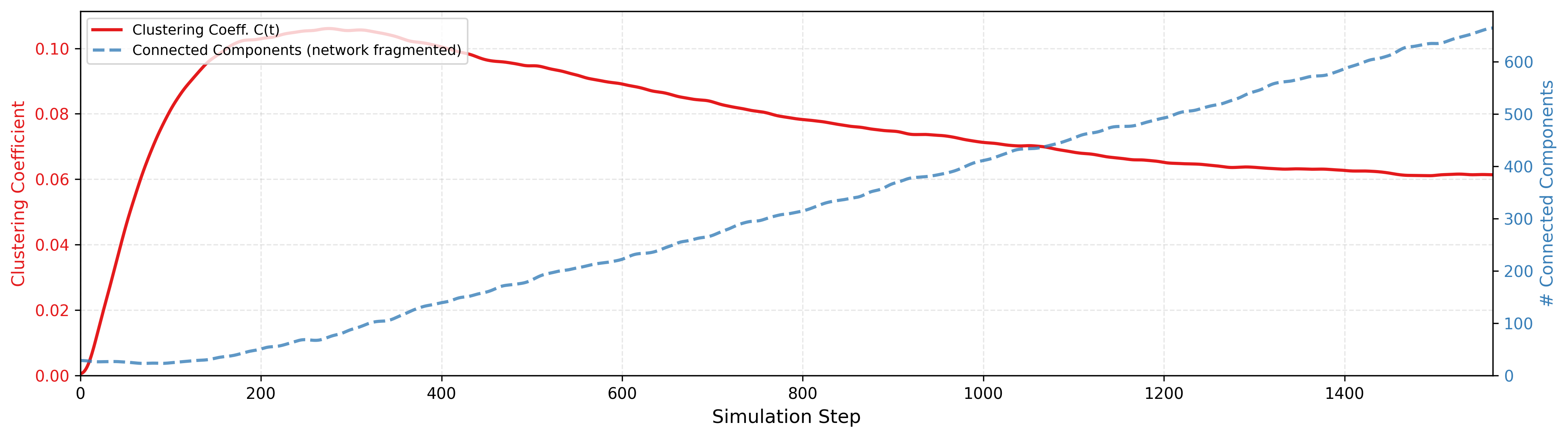}
        \caption{Global clustering coefficient $C(t)$ (red) and number of connected 
        components (blue, right axis), reflecting a network that remains fragmented 
        throughout the simulation.}
        \label{fig:dynamics_baseline_c}
    \end{subfigure}
    \caption{Longitudinal dynamics of the simulation under the pure greedy baseline 
    (0\% strategic agents).}
    \label{fig:dynamics_baseline}
\end{figure}

Figure~\ref{fig:dynamics_baseline} presents the structural baseline of the simulation under the pure greedy regime (0\% strategic agents). Specifically, Figure~\ref{fig:dynamics_baseline_a} shows that paper production reaches a stable pipeline relatively quickly (around 100 simulation steps), with cumulative completions growing at roughly $86.8$ papers per step early in the simulation and $93.6$ in the late phase. Figure~\ref{fig:dynamics_baseline_b} documents one of the more striking features of the baseline: the Gini coefficient of cumulative utilities falls sharply from $0.994$ at the outset to $0.212$ by the simulation's end. This outcome can be explained by the fact that nearly all agents take the earliest steps have completed few or no papers, so the distribution is dominated by the handful of authors who happen to finish first. Later, as publications accumulate more broadly, inequality converges to a stable level. Figure~\ref{fig:dynamics_baseline_c} reveals that the friendship network remains fragmented throughout, settling into $666$ connected components with a global clustering coefficient of $C = 0.061$, which is well above the Erd\H{o}s--R\'{e}nyi expectation of $C_{\mathrm{ER}} \approx 0.002$ at the same density \cite{erdos1959random}. The absence of a single connected component is itself a meaningful result, as it suggests that academic collaboration naturally organizes into semi-isolated local clusters rather than a single integrated community, a pattern broadly consistent with empirical co-authorship networks~\cite{newman2001structure}.

\begin{figure}[!ht]
    \centering
    \includegraphics[width=0.95\linewidth]{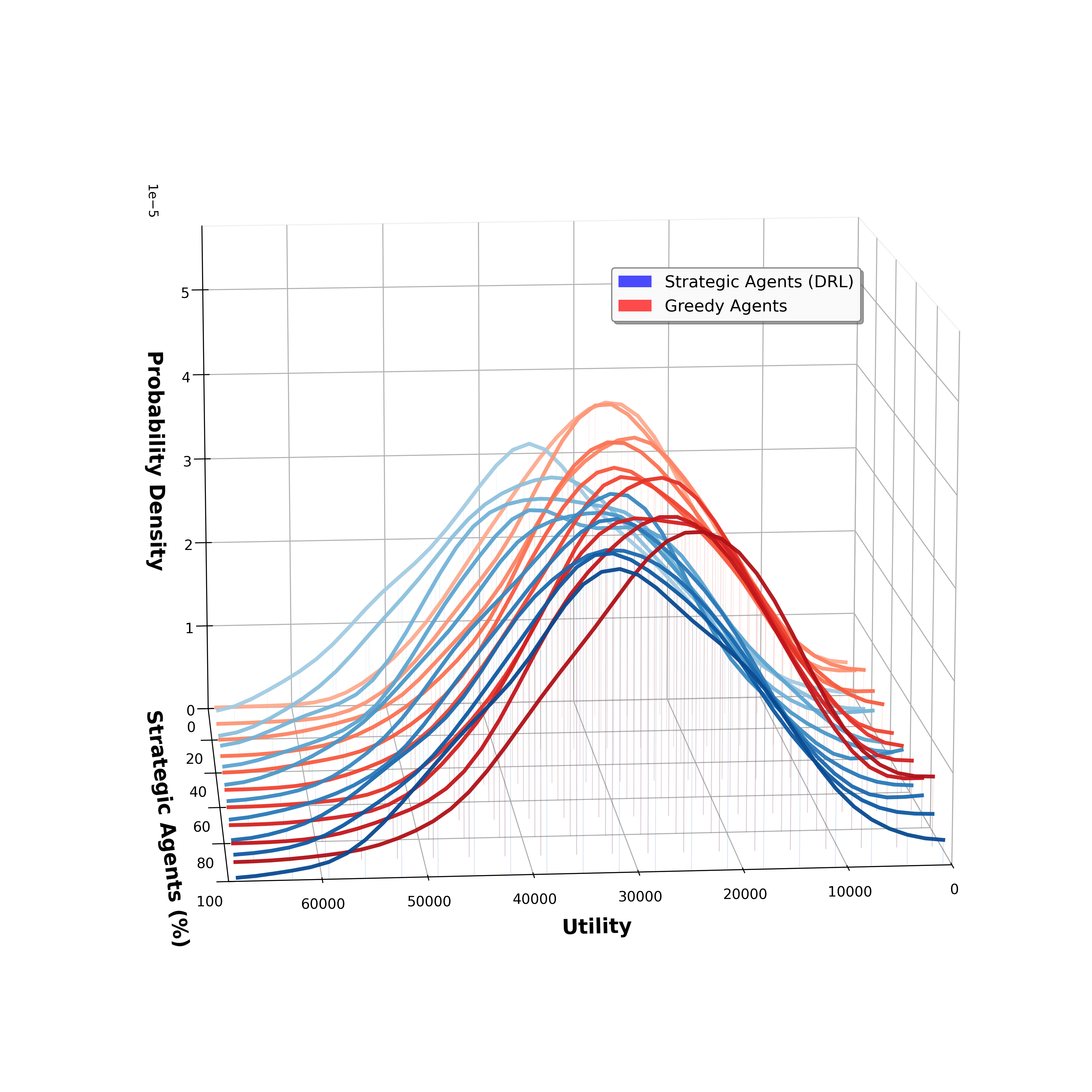}
    \caption{Utility distributions for strategic (blue) and greedy (red) agents across varying population compositions. Each curve represents the probability density of cumulative utilities at a fixed strategic-agent percentage (0\% to 100\% in 10\% increments.}
    \label{fig:utility_distribution_mixed}
\end{figure}

Figure~\ref{fig:utility_distribution_mixed} shows the utility distributions 
for strategic (blue) and greedy (red) agents across all population compositions. In every configuration, the strategic distributions sit to the right of their greedy counterparts - strategic agents consistently earn more than their myopic peers in the same environment. The gap is largest when strategic agents are rare: at 10\% strategic prevalence, the mean utility of strategic agents reaches $35{,}010$ against $26{,}765$ for greedy agents, a relative premium of $30.8\%$. As strategic density grows, this premium contracts, falling to $20.7\%$ at 80\% prevalence. Notably, the mean utility of greedy agents also declines as more strategic agents enter the population, dropping from $27{,}396$ in the pure-greedy baseline to $25{,}370$ at 90\% strategic density. In other words, greedy agents are made worse off not only 
relative to their strategic peers but in absolute terms compared to a world without strategic agents at all. The exact statistical properties of these distributions are provided in the Appendix, Table~\ref{tab:mixed_results}.

Table~\ref{tab:core_outcomes} summarizes system-level outcomes as the fraction of strategic agents varies from 0\% to 100\%. The table contains the following propertoes: \textit{Ultimata/Paper}: mean number of ultimatums raised per paper, broken down by accepted (Acc.), withdrawn (With.), and terminated (Term.) outcomes. \textit{Paper Rates}: paper completion and destruction  rates. \textit{Pap/Ag}: mean number of completed papers per agent. $\mu_U$\,($\sigma_U$): mean and standard deviation of cumulative utilities for strategic, greedy, and combined populations. \textit{Gini}: Gini coefficient of cumulative utilities for each subpopulation. Importantly, the total number of ultimatums raised per paper changes is remind almost constant, from $0.937$ in the pure-greedy case to $1.019$ in the fully strategic population. What changes dramatically is what happens after an ultimatum fails: the termination rate per paper falls from $0.120$ under the pure-greedy regime to $0.000$ when all agents are strategic, while the withdrawal rate rises from $0.391$ to $0.694$. Moreover, as terminations disappear, paper throughput improves substantially as indicated by the completion rate, which rises from $0.853$ to $0.970$, and the mean number of completed papers per agent grows from $15.205$ to $16.857$. Notably, the Gini coefficient of cumulative utilities stays near $0.21$--$0.23$ across all compositions and for both subpopulations, suggesting that the shift toward strategic behavior raises the overall level of academic output without altering its distribution.

\begin{table}[!ht]
\centering
\caption{System-level indicators of ultimatum behavior, paper throughput, and 
inequality across all population compositions. }
\label{tab:core_outcomes}
\resizebox{\textwidth}{!}{%
\begin{tabular}{c|cccc|cc|c|ccc|ccc|ccc}
\hline
 & \multicolumn{4}{c|}{\textbf{Ultimata / Paper}} 
 & \multicolumn{2}{c|}{\textbf{Paper Rates}} & 
 & \multicolumn{3}{c|}{\textbf{$\mu_{U}$ ($\sigma_{U}$)}} 
 & \multicolumn{3}{c|}{\textbf{Advantage}} 
 & \multicolumn{3}{c}{\textbf{Gini}} \\
\textbf{Strat.\%} & Total & Acc. & With. & Term. 
 & Compl. & Destr. & Pap/Ag 
 & Strat. & Greedy & All 
 & Abs. & Rel.\% & 
 & All & Strat. & Greedy \\
\hline
0 & 0.937 & 0.427 & 0.391 & 0.120 & 0.853 & 0.120 & 15.205 
  & --- & 27{,}396 (10{,}841) & 27{,}396 (10{,}841) 
  & --- & --- & & 0.212 & --- & 0.212 \\
10 & 0.941 & 0.388 & 0.449 & 0.105 & 0.869 & 0.105 & 15.301 
   & 35{,}010 (13{,}215) & 26{,}765 (10{,}875) & 27{,}589 (11{,}403) 
   & 8{,}246 & 30.8 & & 0.222 & 0.208 & 0.218 \\
20 & 0.947 & 0.362 & 0.492 & 0.093 & 0.880 & 0.093 & 15.425 
   & 33{,}455 (12{,}280) & 26{,}457 (10{,}840) & 27{,}856 (11{,}489) 
   & 6{,}998 & 26.5 & & 0.223 & 0.203 & 0.221 \\
30 & 0.960 & 0.346 & 0.533 & 0.081 & 0.892 & 0.081 & 15.677 
   & 33{,}204 (12{,}565) & 26{,}296 (10{,}797) & 28{,}369 (11{,}789) 
   & 6{,}908 & 26.3 & & 0.224 & 0.205 & 0.221 \\
40 & 0.976 & 0.339 & 0.568 & 0.069 & 0.903 & 0.069 & 15.758 
   & 32{,}241 (12{,}035) & 26{,}123 (10{,}857) & 28{,}570 (11{,}732) 
   & 6{,}118 & 23.4 & & 0.223 & 0.205 & 0.224 \\
50 & 0.985 & 0.331 & 0.596 & 0.057 & 0.914 & 0.057 & 15.950 
   & 32{,}142 (12{,}320) & 25{,}786 (10{,}938) & 28{,}964 (12{,}075) 
   & 6{,}356 & 24.6 & & 0.227 & 0.210 & 0.228 \\
60 & 0.987 & 0.332 & 0.610 & 0.045 & 0.927 & 0.045 & 16.118 
   & 31{,}691 (11{,}946) & 25{,}663 (10{,}638) & 29{,}280 (11{,}816) 
   & 6{,}028 & 23.5 & & 0.221 & 0.208 & 0.225 \\
70 & 0.998 & 0.327 & 0.638 & 0.033 & 0.938 & 0.033 & 16.367 
   & 31{,}499 (12{,}131) & 25{,}765 (10{,}641) & 29{,}779 (11{,}995) 
   & 5{,}733 & 22.3 & & 0.219 & 0.211 & 0.222 \\
80 & 1.005 & 0.331 & 0.653 & 0.021 & 0.950 & 0.021 & 16.560 
   & 31{,}211 (11{,}714) & 25{,}858 (10{,}749) & 30{,}140 (11{,}725) 
   & 5{,}353 & 20.7 & & 0.213 & 0.207 & 0.223 \\
90 & 1.019 & 0.331 & 0.678 & 0.010 & 0.960 & 0.010 & 16.660 
   & 30{,}974 (11{,}917) & 25{,}370 (10{,}754) & 30{,}413 (11{,}925) 
   & 5{,}604 & 22.1 & & 0.216 & 0.212 & 0.227 \\
100 & 1.019 & 0.325 & 0.694 & 0.000 & 0.970 & 0.000 & 16.857 
    & 30{,}759 (11{,}875) & --- & 30{,}759 (11{,}875) 
    & --- & --- & & 0.211 & 0.211 & --- \\
\hline
\end{tabular}%
}
\end{table}

\begin{figure}[H]
    \centering
    \begin{subfigure}[t]{0.82\textwidth}
        \centering
        \includegraphics[width=\textwidth]{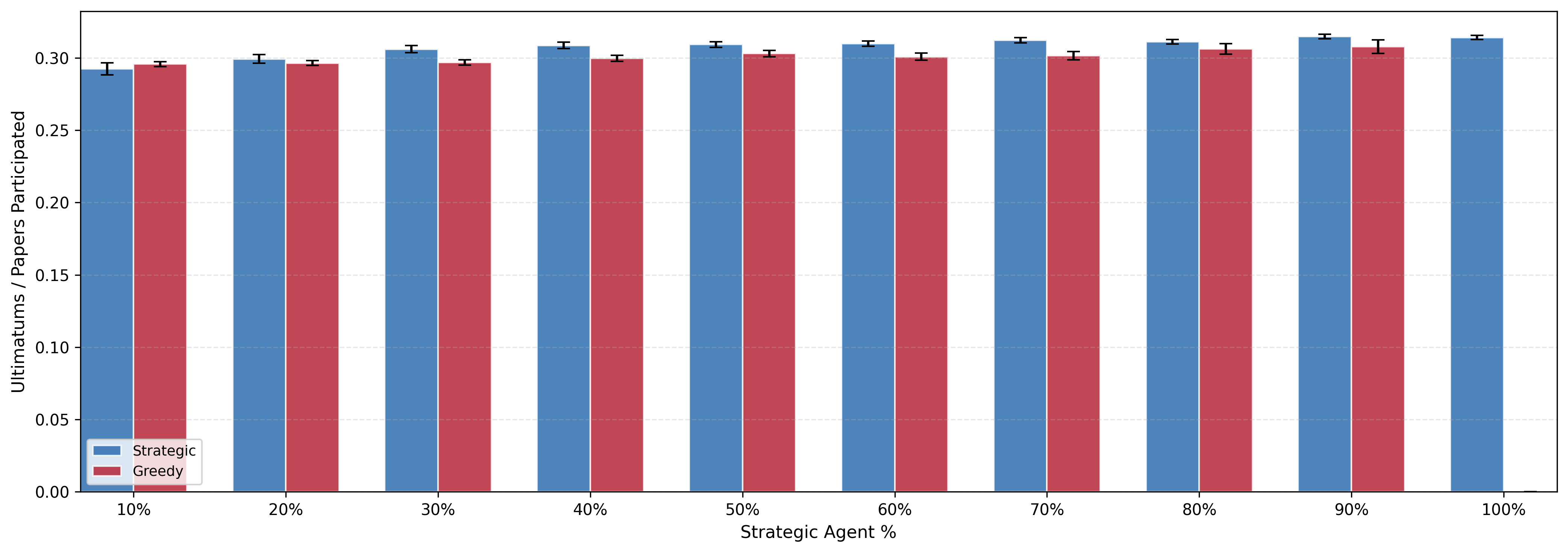}
        \caption{Per-agent initiation rate (ultimatums raised per paper participated) 
        for strategic (blue) and greedy (red) agents, with $\pm 1$ SE error bars; 
        rates are nearly identical across all compositions.}
        \label{fig:ult_a}
    \end{subfigure}
    \\[1ex]
    \begin{subfigure}[t]{0.82\textwidth}
        \centering
        \includegraphics[width=\textwidth]{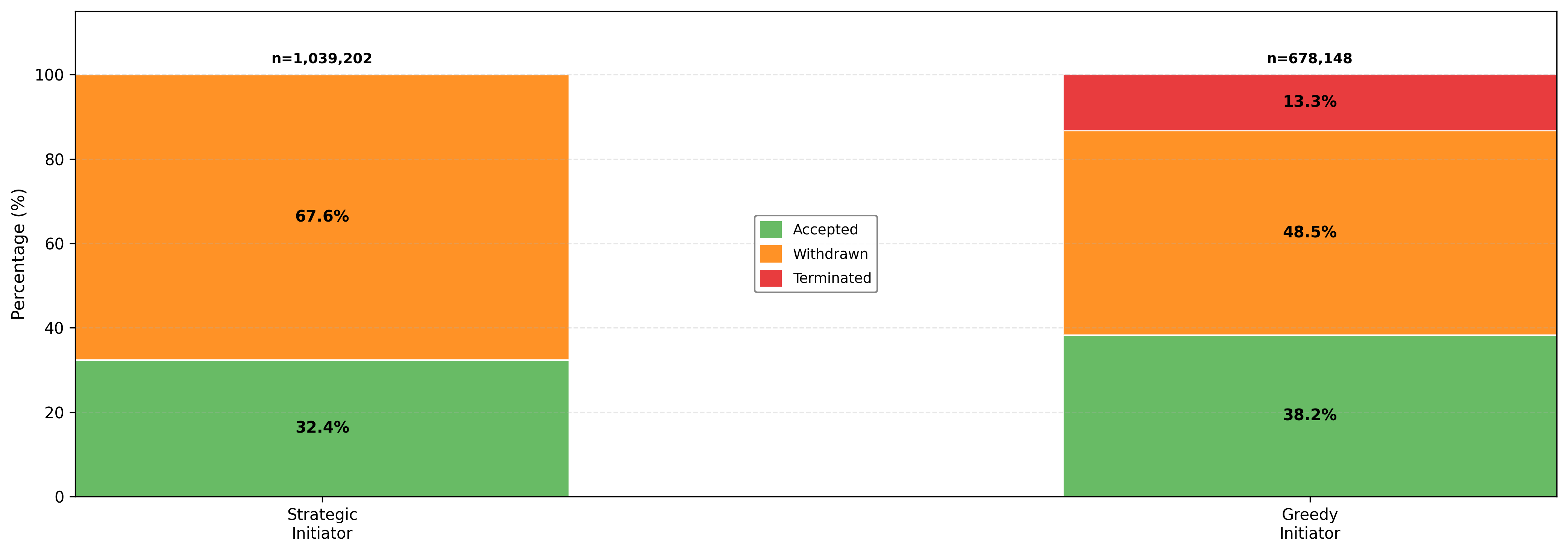}
        \caption{Outcome composition (accepted, withdrawn, terminated) by initiator 
        type, aggregated across all mixed-population configurations.}
        \label{fig:ult_b}
    \end{subfigure}
    \\[1ex]
    \begin{subfigure}[t]{0.82\textwidth}
        \centering
        \includegraphics[width=\textwidth]{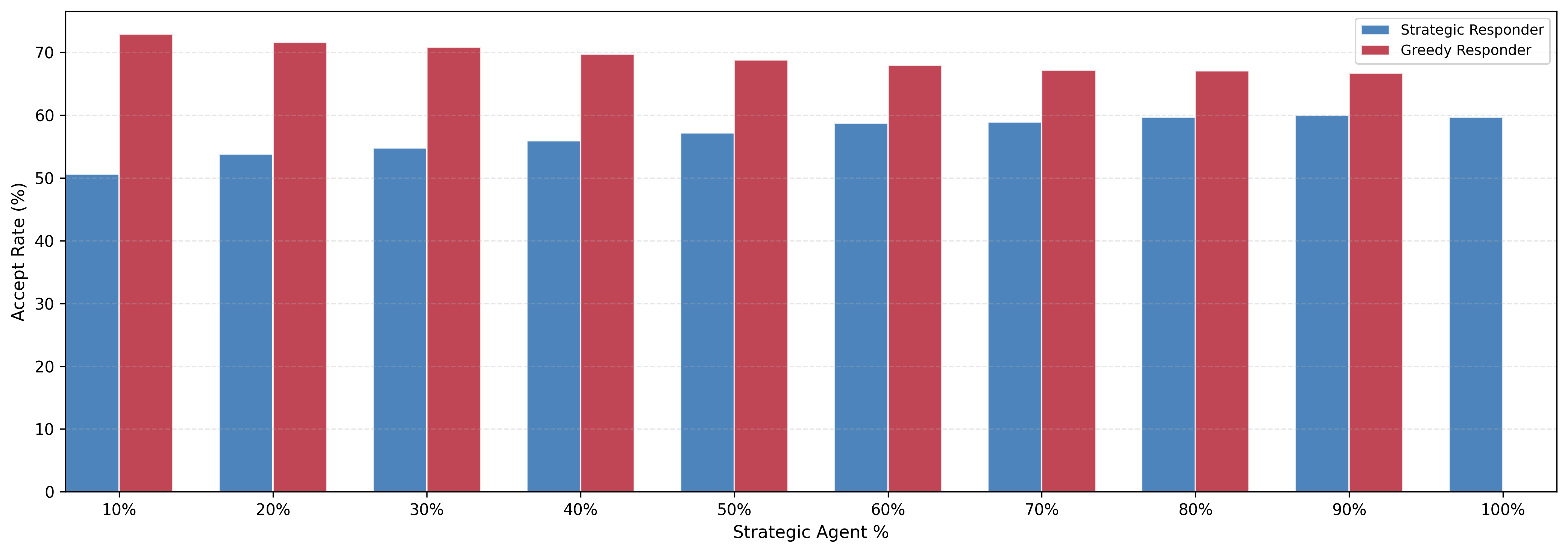}
        \caption{Acceptance rate when a strategic versus a greedy agent acts as 
        responder to an incoming ultimatum, across all population compositions.}
        \label{fig:ult_c}
    \end{subfigure}
    \\[1ex]
    \begin{subfigure}[t]{0.82\textwidth}
        \centering
        \includegraphics[width=\textwidth]{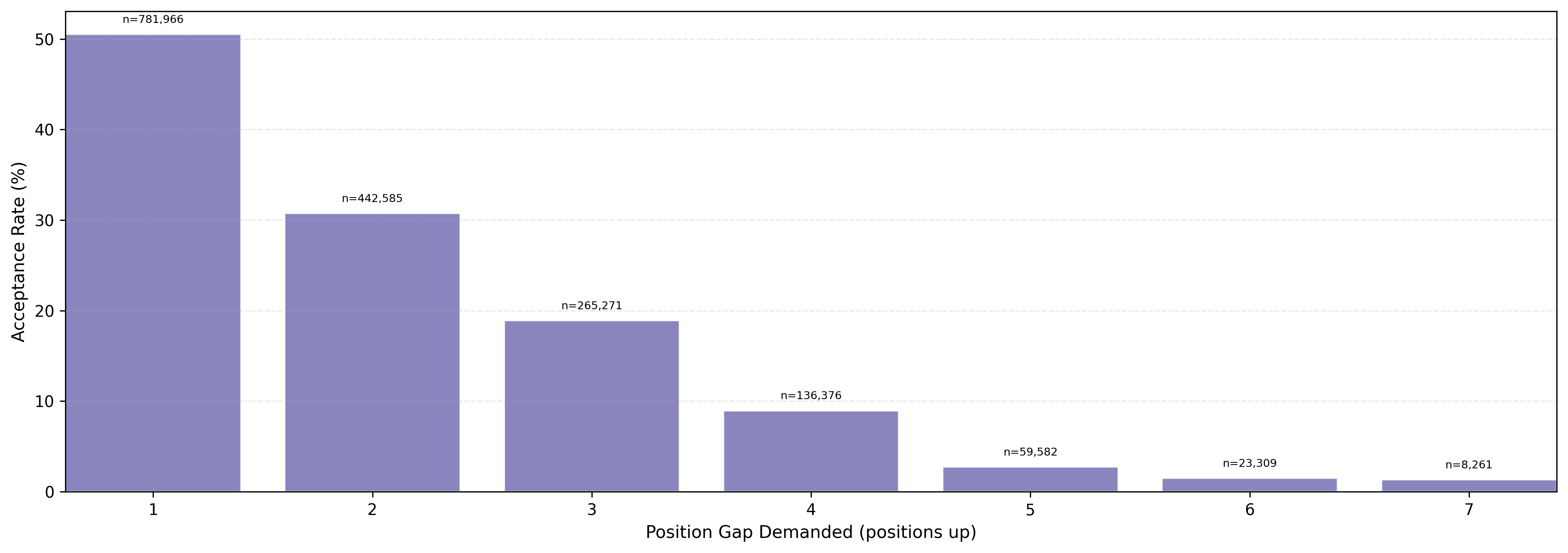}
        \caption{Acceptance rate as a function of the position gap demanded 
        (1--7 positions), aggregated across all configurations.}
        \label{fig:ult_d}
    \end{subfigure}
    \caption{Ultimatum behavior by agent type across population compositions.}
    \label{fig:ultimatum_behavior}
\end{figure}

Figure~\ref{fig:ultimatum_behavior} outlines ultimatum behavior by agent type. Namely, Figure~\ref{fig:ult_a} shows that strategic and greedy agents raise ultimatums at essentially the same rate, ranging between $0.29$ and $0.31$ per paper participated, ruling out the possibility that the DRL policy simply learns to avoid raising demands altogether. Figure~\ref{fig:ult_b} shows that among the $1{,}039{,}202$ ultimatums raised by strategic initiators, the termination rate is zero without exception — every failed demand ends in voluntary withdrawal. Among greedy initiators, by contrast, roughly one in eight ultimatums destroys the manuscript. Figure~\ref{fig:ult_c} shows that strategic agents acting as responders accept incoming ultimatums at noticeably lower rates than greedy responders ($51$--$60\%$ versus $67$--$73\%$), suggesting that the DRL policy also learns that accepting others' demands costs more in positional utility than it saves in reputational terms. Figure~\ref{fig:ult_d} shows that acceptance rates fall steeply with the size of the demanded positional jump, from around $51\%$ for a single-step improvement to below $11\%$ for gaps of four or more positions.
The exact statistical properties of these distributions are provided in the Appendix, Table~\ref{tab:ultimatum_dynamics}.

\begin{figure}[H]
    \centering
    \begin{subfigure}[t]{0.85\textwidth}
        \centering
        \includegraphics[width=\textwidth]{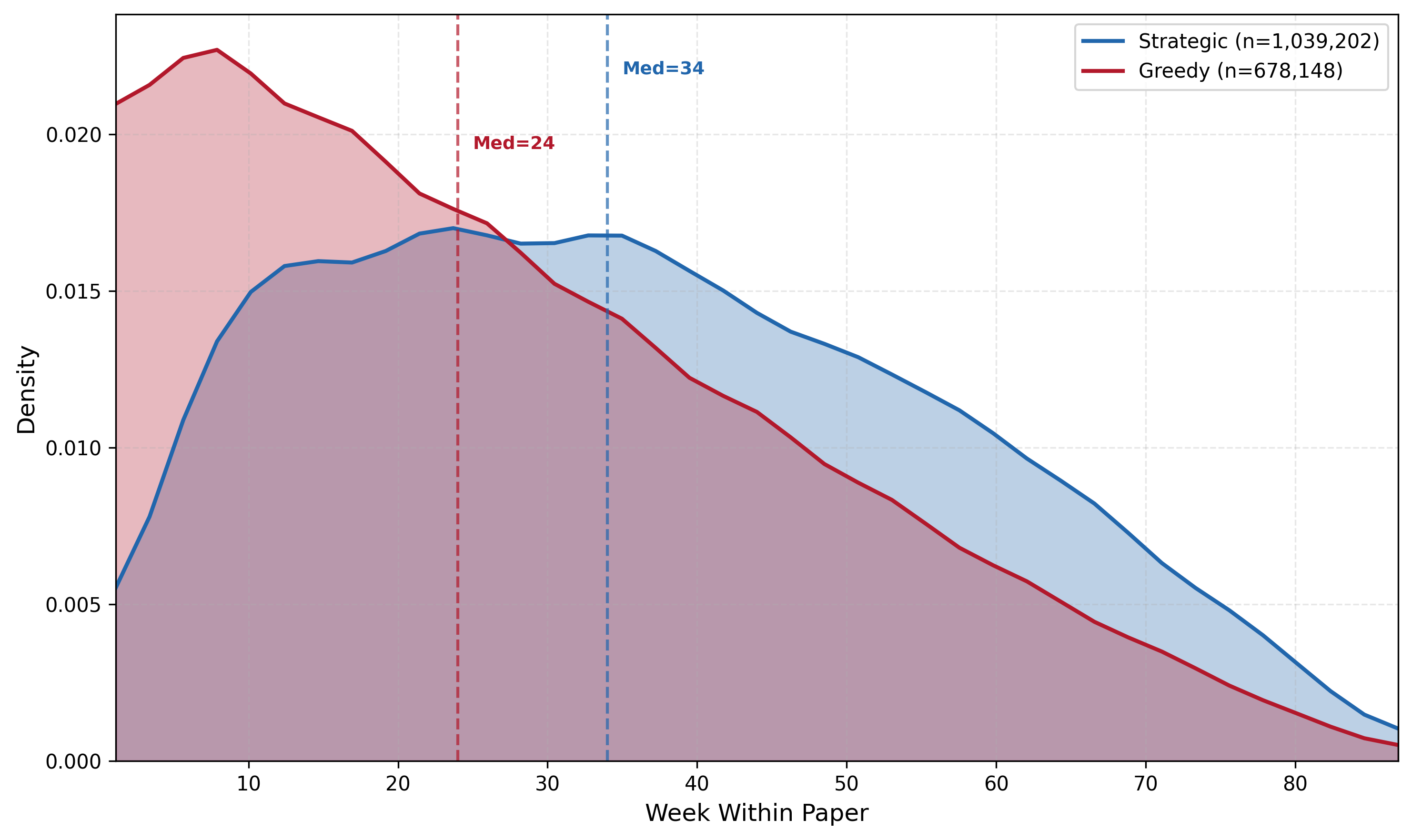}
        \caption{Density of the week at which ultimatums are raised, shown separately 
        for strategic (blue, $n=1{,}039{,}202$) and greedy (red, $n=678{,}148$) 
        initiators; dashed vertical lines indicate the respective medians 
        (Med$=34$ and Med$=24$).}
        \label{fig:timing_kde}
    \end{subfigure}
    \\[1ex]
    \begin{subfigure}[t]{0.85\textwidth}
        \centering
        \includegraphics[width=\textwidth]{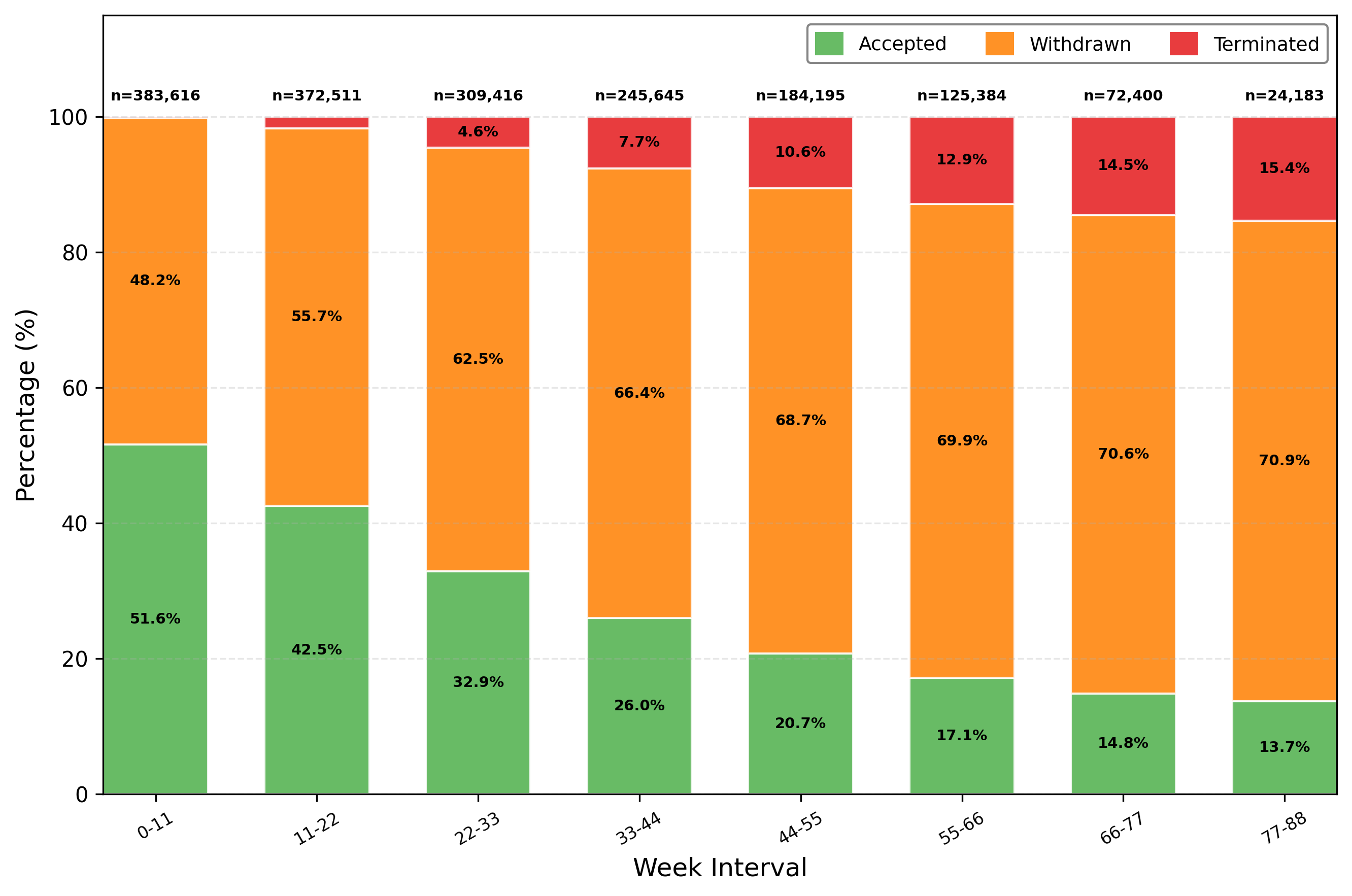}
        \caption{Outcome composition (accepted, withdrawn, terminated) as a function 
        of the week interval in which the ultimatum was raised, aggregated across 
        all mixed-population configurations.}
        \label{fig:timing_outcome}
    \end{subfigure}
    \caption{Temporal dynamics of ultimatums within the paper lifecycle, aggregated 
    across all mixed-population configurations.}
    \label{fig:timing}
\end{figure}

Figure~\ref{fig:timing} examines the timing of ultimatums within the paper lifecycle. Figure~\ref{fig:timing_kde} shows that strategic and greedy agents differ in when they raise ultimatums: while greedy agents peak sharply in the earliest weeks (Med$=24$), strategic agents exhibit a flatter, more delayed distribution with a median of $34$ weeks, suggesting that the DRL policy learns to wait for more favorable conditions before issuing demands. Figure~\ref{fig:timing_outcome} shows why timing nevertheless matters. Ultimatums raised early in a project, when sunk costs are low, are accepted at around $51\%$ and result in termination only $2.4\%$ of the time. By the final weeks of a project, acceptance rates have fallen to roughly $14\%$ and termination rates have risen sharply; unless strategic agents are present, in which case the insistence decision is suppressed, and destruction rates approach zero regardless of timing.

\begin{figure}[H]
    \centering
    \includegraphics[width=0.99\linewidth]{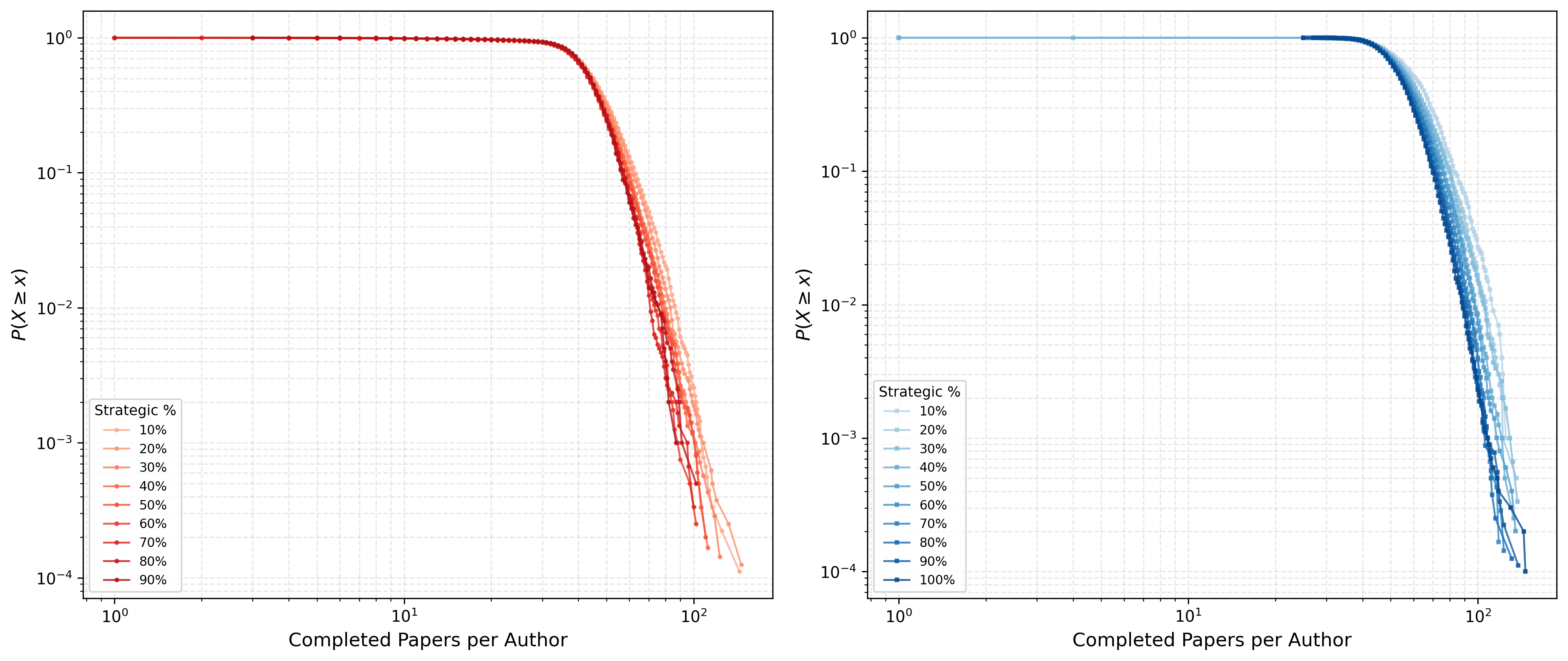}
    \caption{Productivity CCDF by agent type across population compositions. 
    Each panel plots the complementary cumulative distribution function 
    $P(X \geq x)$ of completed papers per author on log--log axes. 
    \textit{Left}: greedy agents across mixed-population configurations 
    (10\%--90\% strategic prevalence); curves are tightly clustered, 
    indicating that greedy productivity is largely invariant to population 
    composition. \textit{Right}: strategic agents across configurations 
    (10\%--100\% strategic prevalence); curves shift leftward and compress 
    as strategic density increases, reflecting the erosion of individual 
    advantage under intensified competition among forward-looking agents. 
    Darker shades correspond to higher fractions of strategic agents.}
    \label{fig:lotka_ccdf_overlay}
\end{figure}

Figure~\ref{fig:lotka_ccdf_overlay} plots the complementary cumulative distribution of completed papers per author on log--log axes for each population composition. The two panels tell qualitatively different stories. For greedy agents, the curves are tightly clustered regardless of how many strategic agents are present; mean output stays between $43.1$ and $45.7$ papers per agent, and the fitted power-law exponent barely alters. Greedy agents are largely passive recipients of the collaboration environment rather than active shapers of it. For strategic agents the picture changes as their numbers grow: mean output falls from $62.0$ papers at 10\% prevalence to 
$54.8$ at 100\%, the power-law tail exponent rises from $9.17$ to $12.81$, and the lognormal $\hat{\sigma}$ contracts from $0.29$ to $0.20$. The highest producers are drawn toward the mean as competition among strategic agents intensifies and the advantages available in a largely myopic environment erode. The top-$10\%$ share declines for both groups as strategic density grows, from $15.3\%$ to $14.1\%$ for strategic agents and from $15.9\%$ to $14.5\%$ for greedy agents. Rather than concentrating output among the most capable agents, a higher prevalence of strategic behavior appears to equalize it. Full distributional statistics are reported in the Appendix, Table~\ref{tab:lotka_stats}.

\section{Discussion}
\label{sec:discussion}
In this study, we extend the single-project co-authorship ultimatum model of 
\cite{lazebnik2023academic} into a repeated, networked game in which researchers collaborate across multiple papers over time, and a subset of agents is trained via DRL to maximize long-run cumulative utility. The model allows us to move beyond the question of whether raising an ultimatum is rational in a single manuscript and ask when such behavior is sustainable (or self-defeating) in an interconnected academic community. 

With respect to \textbf{RQ1}, which asks what the conditions are for local optimal ultimatums with long-term goals. Table \ref{tab:core_outcomes} shows that strategic agents do not raise fewer ultimatums than greedy ones; they raise them at a similar rate. What they learn, instead, is never to insist after rejection, as indicated by Figure \ref{fig:ultimatum_behavior}. Importantly, the decision to issue an ultimatum reflects a short-run 
assessment of positional gain, while the decision to insist or withdraw after rejection is where long-run reasoning enters. The strategic agents effectively decouple these two choices, preserving the option to attempt renegotiation while eliminating the threat of project destruction that gives the ultimatum its leverage in the original model \cite{lazebnik2023academic}. In this context, the timing analysis reinforces this interpretation as ultimatums are most dangerous in the late stages of a project, when sunk costs are 
high, and co-authors are unwilling to accept positional demotions, yet strategic agents suppress insistence regardless of timing, converting what would otherwise be catastrophic late-stage breakdowns into peaceful withdrawals.

With respect to \textbf{RQ2}, Figure \ref{fig:utility_distribution_mixed} confirm that strategic agents benefit at the direct expense of greedy agents. The utility premium of forward-looking agents is largest when they are rare, a 30.8\% advantage at 10\% strategic prevalence, and erodes as their numbers grow, falling to 20.7\% at 80\% prevalence. This is consistent with a competitive externality: when strategic agents are few, they operate in an 
environment populated mainly by myopic partners who can be outmaneuvered without triggering the reputational penalties that a fully strategic population would impose. As their numbers grow, the pool of exploitable partners shrinks and competition among strategic agents intensifies. The more unsettling finding, however, concerns greedy agents: their mean utility declines not only relative to strategic agents but in absolute terms, falling below the pure-greedy baseline as strategic prevalence rises \cite{lazebnik2023academic}.

One result that might seem counterintuitive is the stability of output inequality across all population compositions. The Gini coefficient of cumulative utilities fluctuates narrowly between $0.211$ and $0.227$, as presented in Figure \ref{tab:core_outcomes}, regardless of how many strategic agents are present, and the productivity analysis shows that increasing strategic prevalence compresses rather than stretches the output distribution. The highest producers are drawn toward the mean as competition intensifies, and the top-10\% share declines for both agent types. Rather than concentrating rewards among the most capable or opportunistic agents, the reputational feedback embedded in the friendship network appears to act as an equalizing force: agents who destroy projects accumulate reputational penalties that limit their future collaboration opportunities, preventing 
any single strategy from generating runaway advantage. At the population level, a fully strategic environment achieves what amounts to a Pareto improvement over the pure-greedy baseline without widening the productivity gap between researchers.

This study is not without limitations. First, the model assumes that authors are fully informed about the state of their project and the friendship network. In practice, the project duration can only be estimated, and researchers assess their collaborators' utilities and contributions with considerable noise. Extending the model to partial observability would make the DRL task harder, but would likely produce more conservative strategic 
behavior, as agents would need to hedge against uncertainty in the actions of others. Second, the model abstracts away several features of real academic life, including field-specific authorship norms, institutional hierarchies, and the power differential between senior and junior researchers. These factors are known to shape authorship disputes in practice~\cite{savchenko2024authorship} and their inclusion would allow for a more direct comparison with survey evidence. Third, while the friendship network evolves in response to collaboration outcomes, tie formation is not itself strategic as agents do not deliberately cultivate relationships in anticipation of future projects. Allowing endogenous network formation would connect the present framework to the broader literature on strategic network formation~\cite{jackson1996strategic,carayol2009knowledge} and would raise the question of whether forward-looking agents invest differently in their professional networks than myopic ones. Following these limitations, future work can extend the present model in several directions. One possible extension is to allow agents to differ in their discount factors and risk attitudes, moving beyond the binary strategic-greedy distinction used here. Another is to calibrate the network and paper-formation parameters to empirical co-authorship data from specific disciplines, so that the model's predictions about ultimatum prevalence and inequality can be tested against the survey evidence in~\cite{savchenko2024authorship}. Perhaps most practically, the DRL framework developed here could serve as a testbed for 
evaluating institutional remedies by training the policy under different 
penalty regimes and assessing which mechanism designs reduce destruction rates without requiring implausibly rigid commitments at the outset of a collaboration.

Taken jointly, the results of this study provide a hopeful message about the future of scientific collaboration. While the modern co-authorship system creates clear incentives for conflict over credit, our findings suggest that repeated interaction and reputational feedback can discipline these incentives in ways that benefit both individuals and the broader community. Strategic, forward-looking behavior does not eliminate competition, but it redirects it away from destructive outcomes and toward more sustainable collaboration. As a result, the academic system represented by the model becomes more productive, more stable, and no less equitable. The framework developed here, therefore, contributes both a novel theoretical lens on authorship conflict and a practical simulation environment for exploring policies that could strengthen trust, fairness, and efficiency in collaborative research.

\section*{Declarations}
\subsection*{Funding}
This study received no funding.

\subsection*{Conflicts of interest/Competing interests}
None.

\subsection*{Code and Data availability}
All the code and data used for this study are freely available in \url{https://github.com/AmitBengiat/risky-game-2}.

\subsection*{Acknowledgments}
The authors wish to thank Ariel Rosenfeld for the brainstorming during the development of this study. 

\subsection*{Author Contribution}
Amit Bengiat: Methodology, Software, Formal analysis, Investigation, Resources, Data Curation, Writing - Original Draft, Writing - Review \& Editing, Visualization. \\ Teddy Lazebnik: Conceptualization, Methodology, Validation, Formal analysis, Investigation, Writing - Original Draft, Writing - Review \& Editing, Visualization, Supervision. 
 
\bibliography{biblio}
\bibliographystyle{unsrt}

\section*{Appendix}

\subsection*{DRL model formalization}
In this section, we formally outline the proposed DRL's model state representation, action space, deep neural network architecture, and training and inference procedures.  

\subsubsection*{State representation}
We represent the state of author $p_i$ at time $t$ as \(s_t^i = \bigl(x_{t,\text{paper}}^i,\; x_{t,\text{agent}}^i,\; x_{t,\text{network}}^i\bigr)\), which concatenates three groups of features: Paper–level context ($x_{t,\text{paper}}^i$) which encode the immediate stakes in the currently active paper $m$ for which $p_i \in K_m^{(t)}$; Agent–level history (\(x_{t,\text{agent}}^i\)) which summarize the author's career and risk profile; and network context ($x_{t,\text{network}}^i$) which summarize the author’s local position and reputation in the friendship network $G_{\text{friendship}}^{(t)}$.

Formally, $x_{t,\text{paper}}^i$ includes \(|K_m^{(t)}|,\quad \frac{|K_m^{(t)}|}{K_{\max}}\), where $K_{\max}$ is the maximal possible clique size; contribution profile in the form \(\text{rel\_contrib}_i = \text{contrib}_{i,m}/\sum_{p_j \in K_m^{(t)} \text{contrib}_{j,m}}\), capturing $p_i$'s relative effort and inequality of efforts; authorship position \(j_{i,m},\quad \frac{j_{i,m}}{|K_m^{(t)}|},\quad \mathbb{I}\{j_{i,m}=1\},\quad \mathbb{I}\{j_{i,m}=|K_m^{(t)}|\}\); paper utility parameters as a scalar paper value $u^0_{i,m}$ and a low–dimensional parameterization of the position–dependent utility $u^1_{i,m}(j)$ such that $(\eta_{i,m},\xi_{i,m})$ if \(u^1_{i,m}(j) = 1 - \eta_{i,m}\, j + \xi_{i,m}\); temporal status \(\frac{\tau}{T_m},\quad \frac{T_m - \tau}{T_m}\),  where $\tau$ is the internal week and $T_m$ the (random) completion time, encoding how close the paper is to completion; myopic gain from raising an ultimatum \(\Delta u_{i,m}^{\text{myopic}} = u^0_{i,m}\bigl(u^1_{i,m}(j'_i) - u^1_{i,m}(j_{i,m})\bigr)\), where $j'_i < j_{i,m}$ is the best position $p_i$ could demand in the ultimatum subgame, computed under the existing game–theoretic rules which makes the greedy solution visible to the DRL agent.

\(x_{t,\text{agent}}^i\) includes discounted cumulative utility so far \(U_i^{\text{past}} = \sum_{m \in M_i,\ t_m < t}          \frac{u^0_{i,m} u^1_{i,m}(j_{i,m})}{(1+\rho)^{t_m}}\), where $\rho$ is the economic discount factor and $M_i$ the set of papers involving $p_i$; current workload \( N_i^{\text{active}}(t) = \bigl|\{m : p_i \in K_m^{(t)}\}\bigr|\); ultimatum statistics, \(N_i^{\text{raise}},\quad N_i^{\text{agree}}\), \(N_i^{\text{refuse}},\quad N_i^{\text{pull}}\), and \(N_i^{\text{insist}},\quad N_i^{\text{destr}}\) that represent counting, respectively, how many ultimata $p_i$ has raised, agreed to, refused, pulled back from, insisted on, and how many papers involving $p_i$ were destroyed due to their insistence.

$x_{t,\text{network}}^i$ includes degree and weighted degree in the graph \(\deg_i^{(t)} = |\mathcal{N}_i^{(t)}|, \qquad      \text{wdeg}_i^{(t)} = \sum_{p_j \in \mathcal{N}_i^{(t)}} w_{ij}^{(t)}\), where $\mathcal{N}_i^{(t)}$ is the set of neighbours of $p_i$ and $w_{ij}^{(t)}$ the friendship weight; reputation with respect to the co–authors \(\bar{w}_{\text{coauth}}^{(t)}(i) = \frac{1}{|K_m^{(t)}|-1} \sum_{p_j \in K_m^{(t)}\setminus\{p_i\}} w_{ij}^{(t)}\); and path–based connectivity which is a coarse summary of $p_i$'s connectivity to the rest of the network \(\bar{d}_i^{(t)} = \frac{1}{|\tilde{\mathcal{N}}_i^{(t)}|} \sum_{p_j \in \tilde{\mathcal{N}}_i^{(t)}} d_G^{(t)}(p_i,p_j)\), where $\tilde{\mathcal{N}}_i^{(t)}$ is a sampled two–hop ego network and $d_G^{(t)}$ is a path–strength measure.

Collectively, these features allow the DRL agent to assess the trade–off between potential positional gains on the current paper and long–term reputational costs induced by altering or breaking collaborations.

\subsubsection*{Action space}
Each author may face three conceptually distinct ultimatum–related decision types: raising an ultimatum (initiator decision), agreeing or refusing another's ultimatum (responder decision), and pulling back or insisting after refusal (post–rejection decision). The \say{raising an ultimatum} decision is available when the rules of the manuscript subgame allow $p_i$ to initiate an ultimatum on the current paper. If $a_t^{\text{raise}}=1$, the environment uses the same game–theoretic rule as in the original model to select the demanded position $j'_i<j_{i,m}$ (e.g.\ the myopically optimal one). The \say{agreeing or refusing another's ultimatum} decision is available when a co–author has raised an ultimatum whose implementation requires $p_i$'s consent. The joint pattern of responses from all affected co–authors determines whether the ultimatum is implemented, fails, or triggers a later pull–back/insist decision for the initiator. The \say{pulling back or insisting after refusal} decision is available when $p_i$ has raised an ultimatum and at least one required co–author has refused. Under the base model, insisting leads to destruction of the paper and associated reputational penalties for $p_i$. At any given time $t$ and for a given agent $p_i$, at most one of these three decisions is active; the environment provides a decision–type label $d_t^i \in \{\text{raise},\text{agree},\text{pull}\}$ indicating which head to use. The overall policy thus produces a triplet of Bernoulli distributions \(
  \pi_\theta^{\text{raise}}(a^{\text{raise}} \mid s), \;
  \pi_\theta^{\text{agree}}(a^{\text{agree}} \mid s),\;
  \pi_\theta^{\text{pull}}(a^{\text{pull}} \mid s).
\)

\subsubsection*{Architecture}
We implement a multi–head deep Q–network (DQN) \cite{mnih2015human,toral2025accelerating} architecture with parameter sharing across authors and separate output heads for the three decision (action) types.

Shared encoders for state components are used to build separate representations of each part of the state. The paper features are embedded via \(h_{t,\text{paper}}^i = \phi_{\text{paper}}(x_{t,\text{paper}}^i)\), where \(\phi_{\text{paper}}\) is a small multilayer perceptron (MLP) with ReLU activations, for example two or three fully connected layers. In the same spirit, the agent features are mapped through \(h_{t,\text{agent}}^i = \phi_{\text{agent}}(x_{t,\text{agent}}^i)\), using the same general MLP structure, so that both content- and author-specific information are transformed into comparable latent spaces. Finally, the network features are encoded as \(h_{t,\text{network}}^i = \phi_{\text{network}}(x_{t,\text{network}}^i)\), which can again be implemented as an MLP, or, if richer relational structure is important, replaced by a graph neural network (GNN) operating on a two--hop ego graph centered at \(p_i\), thereby allowing the embedding to reflect the local connectivity pattern around the author.

Once these component embeddings are available, they are combined by a shared fusion trunk that produces a single latent representation used downstream for decision making. Concretely, the three vectors are concatenated as \(h_t^i = [h_{t,\text{paper}}^i;\ h_{t,\text{agent}}^i;\ h_{t,\text{network}}^i]\), which preserves the separation of information while presenting it in a unified format. This concatenated embedding is then mapped to a compact latent state through \(z_t^i = f_{\text{trunk}}(h_t^i)\), where \(f_{\text{trunk}}\) is a fusion MLP consisting of one or two fully connected layers with ReLU activations and, if helpful for stability and generalization, optional layer normalization or dropout. Because the encoders and trunk are shared, all authors produce latents in the same coordinate system, making it straightforward to share learning signals across the population.

Decisions are handled via multiple Q-output heads, so that each decision type has its own action-value mapping while still benefiting from the shared representation \(z_t^i\). For each decision type \(d \in \{\text{raise},\text{agree},\text{pull}\}\), a separate linear head outputs two Q-values according to \(q_t^{i,d} = W^{(d)} z_t^i + b^{(d)} \in \mathbb{R}^2\), and actions are scored by \(Q_\theta^{d}(s_t^i,a) = q_t^{i,d}[a]\) for \(a \in \{0,1\}\). When \(p_i\) encounters a particular decision type \(d_t^i\) at time \(t\), only the corresponding head \(Q_\theta^{d_t^i}\) is used for action selection and learning updates. 

All parameters \(\theta\) (the encoders, the fusion trunk, and the heads) are shared across authors \(p_i\), which implements centralized training with decentralized execution: trajectories from all agents train a single network, while each author acts independently from their own \(s_t^i\). 

\subsubsection*{Training and inference procedures}
We formulate the problem as an MDP for each author and train the shared network off–policy using simulated episodes.

We cast the simulator into a Markov decision process by defining the state space as \(\mathcal{S}\), the set of all vectors \(s_t^i\) observed for authors \(p_i\) over time. At each time \(t\) and for each \(p_i\), the environment also provides a decision type \(d_t^i \in \{\text{raise},\text{agree},\text{pull}\}\) (or \say{none} if no decision is required), and the corresponding action is binary as \(a_t^i \in \{0,1\}\), with its interpretation determined by \(d_t^i\). The transition kernel \(P\) is not written in closed form but is induced by the agent-based simulator, which updates contributions, authorship orders, paper statuses, and friendship weights according to the joint actions of all authors and the underlying stochastic rules. Each author \(p_i\) receives a scalar reward via a function \(R_i\) that scores transitions in terms of discounted publication utility and penalties for destructive outcomes, and future returns are discounted using \(\gamma_{\text{RL}} \in (0,1)\) at the chosen macro/micro time scale.

With this MDP specification in place, the reward is designed to emphasize realized publication value while discouraging decisions that destroy accumulated work. When a paper \(m\) involving \(p_i\) is successfully completed at time \(t_m\), we assign a sparse publication reward \(r_{t_m}^i = \frac{u^0_{i,m} u^1_{i,m}(j_{i,m})}{(1+\rho)^{t_m}}\) and set \(r_t^i = 0\) at intermediate times. If instead a paper \(m\) is destroyed because \(p_i\) insists on an ultimatum, we apply a destruction penalty \(r_t^i = -\lambda_{\text{destr}} \cdot \text{contrib}_{i,m}\), where \(\lambda_{\text{destr}} > 0\) amplifies the loss of invested work. Optionally, a small shaping term can be added to encourage preservation of the ego network by updating \(r_t^i \leftarrow r_t^i + \lambda_{\text{deg}}\bigl(\text{wdeg}_i^{(t+1)} - \text{wdeg}_i^{(t)}\bigr)\), where \(\lambda_{\text{deg}}\) controls how much network stability influences the reward.

Training then proceeds off-policy with a multi-head DQN over episodes that simulate full careers on a fixed horizon \(t=0,\dots,T\), starting from an initial friendship network and collaboration cliques drawn as in the base model. For action selection, at each time step \(t\) and each author \(p_i\) with an active decision type \(d_t^i\), we construct \(s_t^i\) and choose \(a_t^i \in \{0,1\}\) using \(\varepsilon\)-greedy exploration: with probability \(\varepsilon\) we act randomly, otherwise we exploit via \(a_t^i = \arg\max_{a \in \{0,1\}} Q_\theta^{d_t^i}(s_t^i,a)\). The simulator applies all authors' actions, returns next states \(s_{t+1}^i\) and rewards \(r_t^i\), and for each active decision we store \((s_t^i, d_t^i, a_t^i, r_t^i, s_{t+1}^i)\) in a shared replay buffer \(\mathcal{B}\). At regular intervals we sample minibatches from \(\mathcal{B}\) and compute Double-DQN targets on the relevant head as \(y = r_t^i + \gamma_{\text{RL}} Q_{\theta^-}^{d_t^i}\Bigl(s_{t+1}^i,\; \arg\max_{a' \in \{0,1\}} Q_\theta^{d_t^i}(s_{t+1}^i,a')\Bigr)\), where \(\theta^-\) are slowly updated target-network parameters, and we optimize the squared Bellman error \(L(\theta) = \mathbb{E}_{(s,d,a,r,s')\sim\mathcal{B}} \bigl(y - Q_\theta^{d}(s,a)\bigr)^2\), updating only the head indexed by \(d\) for each sample using stochastic gradient descent or Adam. Stabilization follows standard practice by maintaining the target network, using experience replay, and optionally annealing \(\varepsilon\) over time; after convergence, we freeze \(\theta\) and set \(\varepsilon \approx 0\) for deployment. In evaluation simulations, whenever an author \(p_i\) faces a decision of type \(d_t^i\), we construct \(s_t^i\), compute \(Q_\theta^{d_t^i}(s_t^i,0)\) and \(Q_\theta^{d_t^i}(s_t^i,1)\), and choose greedily \(a_t^i = \arg\max_{a \in \{0,1\}} Q_\theta^{d_t^i}(s_t^i,a)\), interpreting the chosen action as raise/not raise, agree/refuse, or pull back/insist depending on \(d_t^i\). In this regime, an author may strategically avoid ultimata that look attractive in the short run (high \(\Delta u_{i,m}^{\text{myopic}}\)) but are harmful in expectation for long-run utility due to reputational damage and network fragmentation, a behavior that the greedy baseline cannot capture.

\subsection*{Deep reinforcement learning training procedure and convergence analysis}
We trained the proposed DRL over 500 episodes, each corresponding to a full simulation run of $T = 1565$ time steps (representing 30 years of research, where each step corresponds to a full week) with a population of $n = 1000$ agents, balancing population size for expressiveness and computational time and memory. To ensure gradual adoption of the strategic policy and to maintain a sufficiently large pool of greedy agents for the DRL agents to learn from, we employed a dynamic conversion schedule: at the start of training, all agents were initialized as greedy, and every 10 episodes, a fixed number $k$ of randomly selected greedy agents were converted to strategic agents (equipped with the partially trained DRL model). The conversion rate $k$ was calibrated so that by episode 400, approximately 80\% of the population had transitioned to strategic mode, with the remaining 20\% serving as a persistent greedy baseline to stabilize the learning signal and prevent overfitting to a purely strategic environment.

The DRL integrator used an $\varepsilon$-greedy exploration strategy \cite{dann2022guarantees} with an exponentially decaying schedule: $\varepsilon_0 = 1.0$ at episode 0, decaying by a factor of 0.9825 per episode to a minimum of $\varepsilon_{\min} = 0.01$ by episode 400, and held constant thereafter. The replay buffer was sized at 100000 transitions, and minibatch updates with a batch size of 32 were performed at regular intervals using the Double DQN target formulation with a target-network update frequency of 1000 simulation steps. The network was optimized via the Adam optimizer \cite{zhang2018improved} with a learning rate of $10^{-4}$, and the RL discount factor was set to $\gamma_{\text{RL}} = 0.99$ to encourage long-horizon credit assignment. Rewards were structured to reflect discounted publication utilities (scaled by $u_{i,m}^0 / 100$ for numerical stability), destruction penalties proportional to wasted effort ($\lambda_{\text{destr}} = 1.0$), and optional reputation-shaping terms ($\lambda_{\text{deg}} = 0.0$ in the reported runs).

\begin{figure}[!ht]
    \centering
    \begin{subfigure}[t]{0.45\textwidth}
        \centering
        \includegraphics[width=\textwidth]{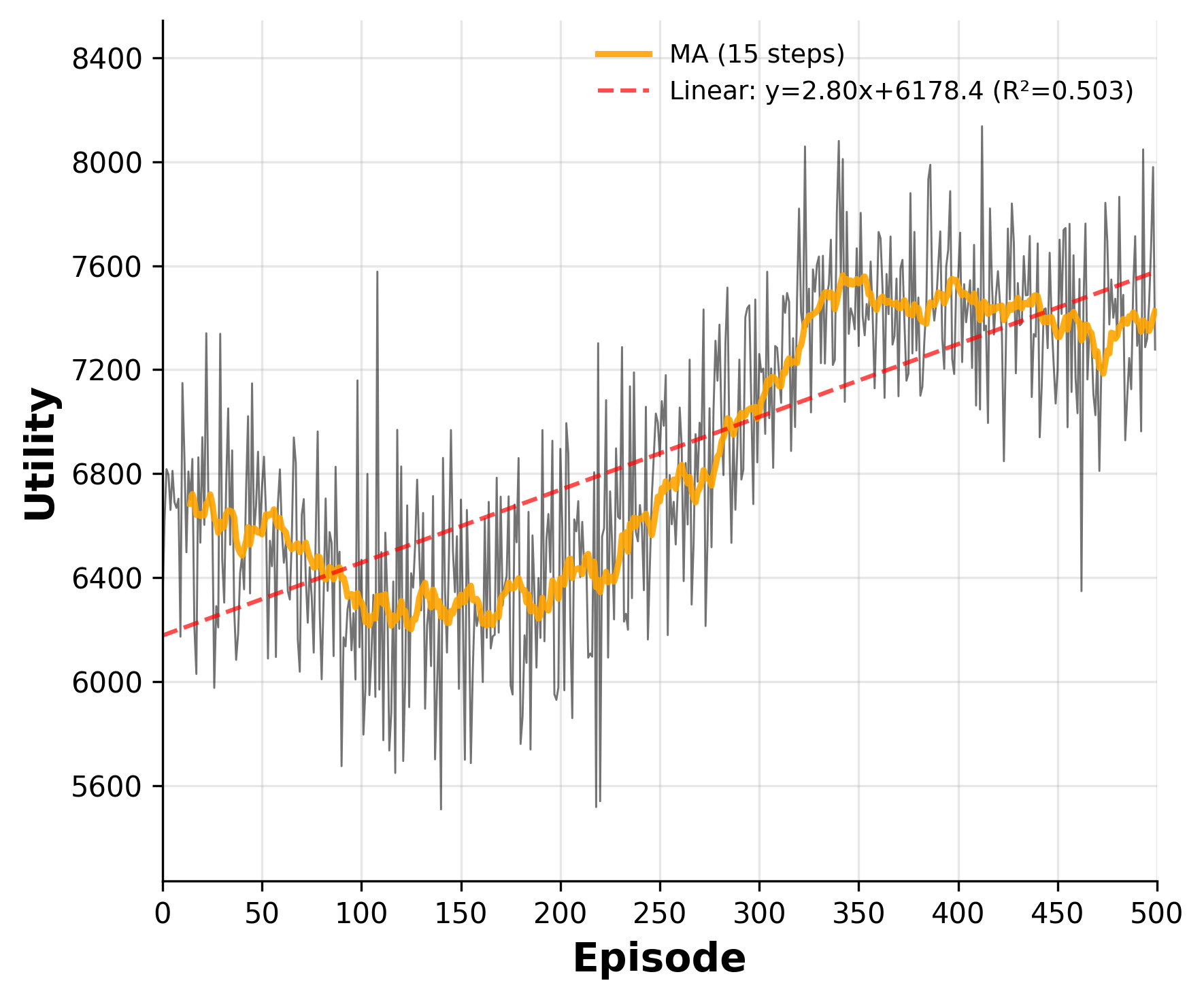}
        \caption{Average utility per episode.}
        \label{fig:subim1}
    \end{subfigure}
    \hfill
    \begin{subfigure}[t]{0.45\textwidth}
        \centering
        \includegraphics[width=\textwidth]{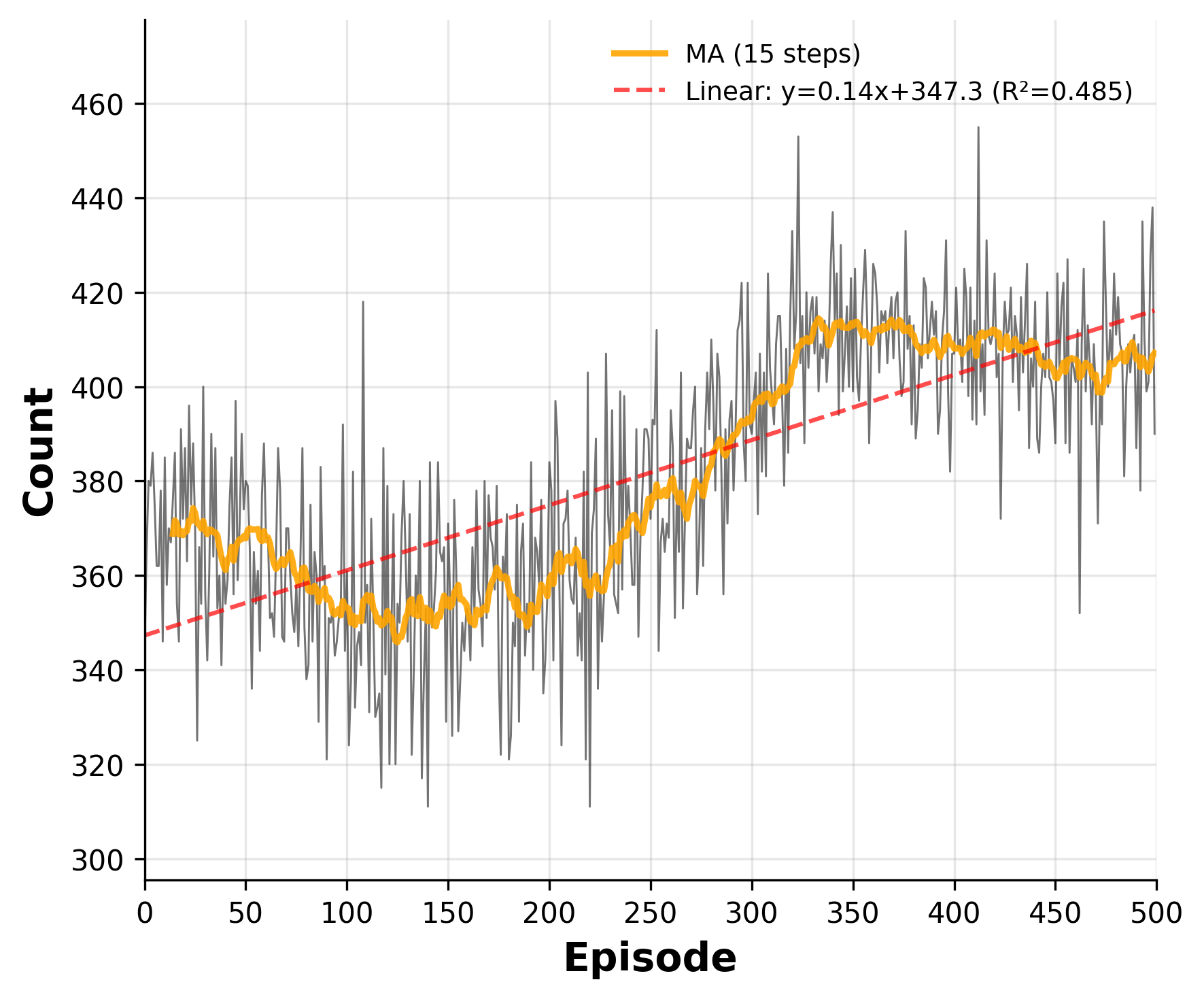}
        \caption{Papers completed per episode.}
        \label{fig:subim2}
    \end{subfigure}
    \\[1ex]
    \begin{subfigure}[t]{0.45\textwidth}
        \centering
        \includegraphics[width=\textwidth]{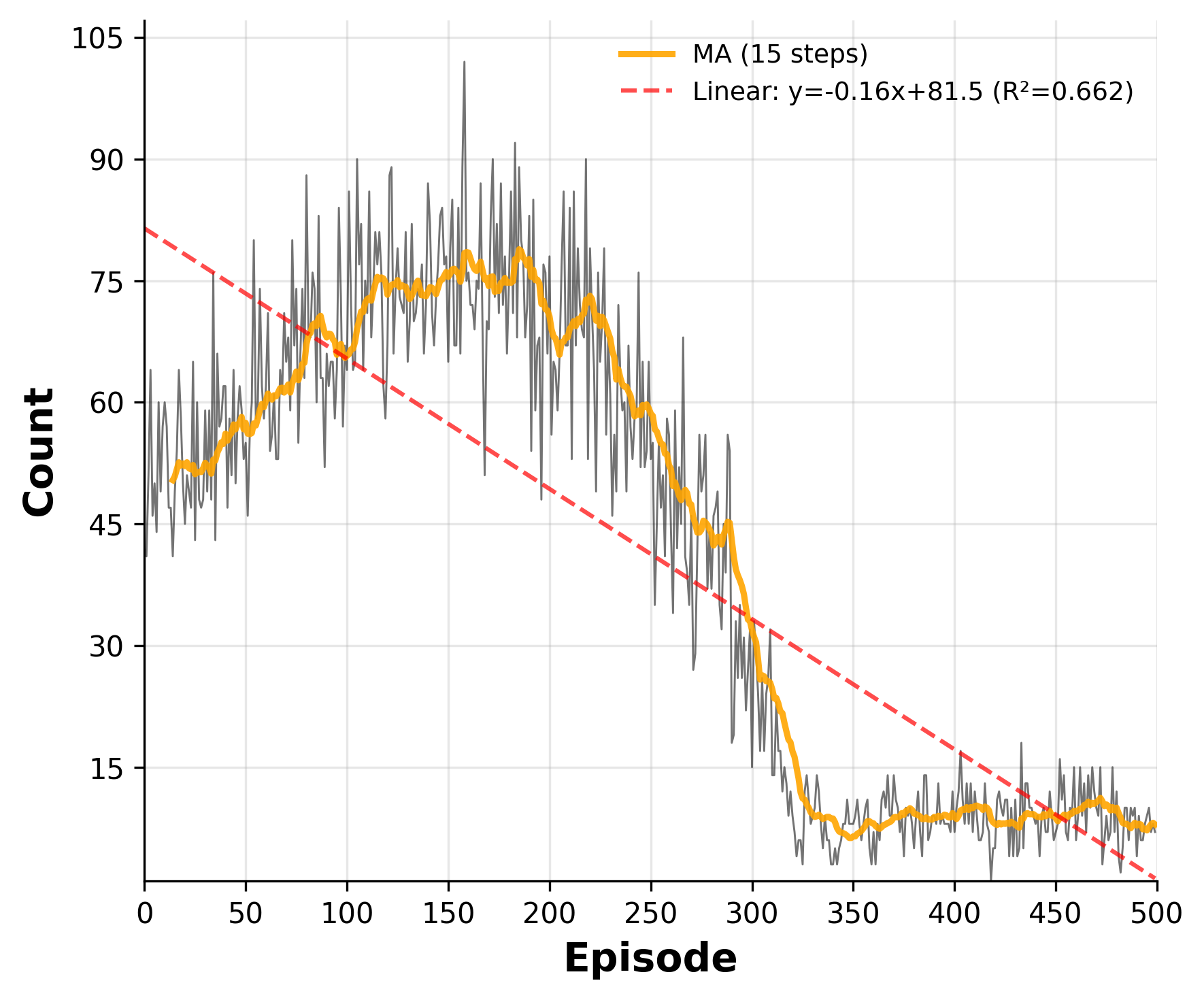}
        \caption{Papers destroyed per episode.}
        \label{fig:subim3}
    \end{subfigure}
    \hfill
    \begin{subfigure}[t]{0.45\textwidth}
        \centering
        \includegraphics[width=\textwidth]{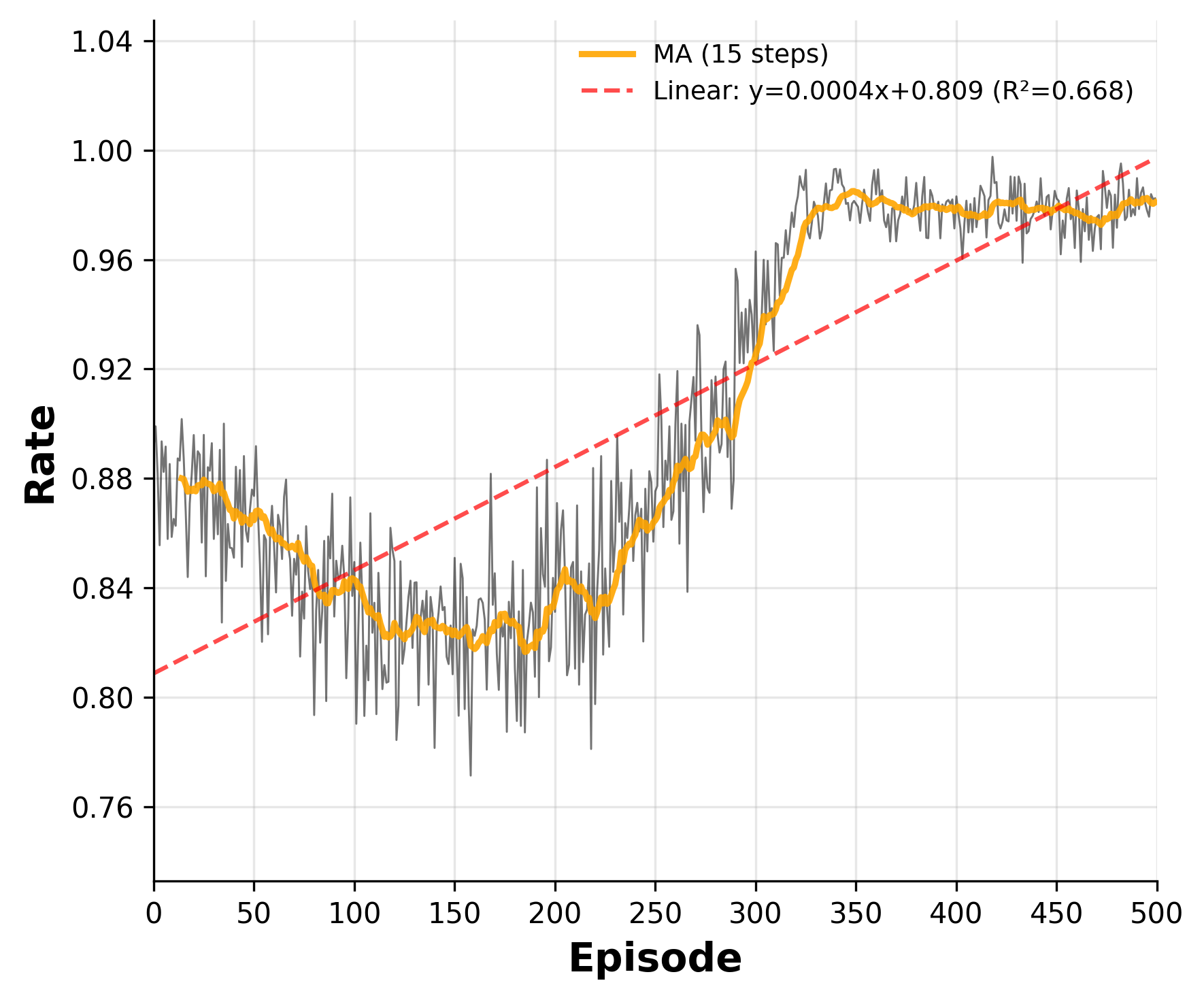}
        \caption{Paper completion rate.}
        \label{fig:subim4}
    \end{subfigure}
    \caption{Training progress of DRL agents over 500 episodes. Each panel shows raw per-episode values (black), a 15-step moving average (orange), and a linear regression trend (red dashed). Average utility increases steadily, paper destructions decline, and the completion rate rises from approximately 0.80 to 0.98.}
    \label{fig:training_progress}
\end{figure}

Figure~\ref{fig:training_progress} summarizes the training dynamics across all 500 episodes. The top-left panel shows the average utility per episode (computed over all agents, with strategic and greedy pooled), which exhibits a clear upward trend from approximately 6200 at episode 0 to over 7400 by episode 500. A 15-episode moving average (orange line) smooths out the high-frequency fluctuations induced by stochasticity in clique formation and paper durations, while a linear regression fit ($y = 2.80x + 6178.4$, $R^2 = 0.503$) quantifies the average rate of improvement. The positive slope indicates that the DRL agents are learning to make decisions that increase overall productivity, consistent with the hypothesis that avoiding destructive ultimata preserves cumulative welfare. The top-right panel tracks the number of completed papers per episode, which similarly rises from around 350 to over 400, with a linear trend line ($y = 0.14x + 347.3$, $R^2 = 0.485$). This increase in throughput suggests that strategic agents are not only optimizing their own utilities but also facilitating more collaborations to reach successful completion, thereby amplifying the total volume of published work.

The most clear evidence of learning appears in the bottom-left panel, which plots the number of papers destroyed per episode. This quantity declines sharply from approximately 80 at the start of training to under 10 by episode 300, and stabilizes near 5--10 for the remainder of the run. A linear fit ($y = -0.16x + 81.5$, $R^2 = 0.662$) captures the robust negative trend, confirming that the DRL policy effectively suppresses the insistence decision in the post-rejection subgame—precisely the behavior that the training objective was designed to penalize via destruction penalties and reputational feedback. The complementary bottom-right panel shows the paper completion rate (completed papers / total papers started), which rises monotonically from 0.82 to nearly 0.99, with the steepest gains occurring between episodes 100 and 300. A linear regression ($y = 0.0004x + 0.809$, $R^2 = 0.668$) provides a conservative estimate of the trend, though the actual trajectory is closer to a saturating exponential. By episode 400, the completion rate has effectively plateaued at 0.99, meaning that fewer than 1\% of initiated papers are terminated due to ultimatum disputes—a dramatic improvement over the baseline greedy regime, in which termination rates can exceed 15\% under comparable network and utility parameters.

The convergence pattern visible in these plots is consistent with the theoretical prediction that, in a repeated networked game, rational forward-looking agents should eventually learn to avoid locally optimal but globally destructive actions. The initial phase (episodes 0--100) is characterized by high exploration ($\varepsilon$ near 1.0) and a low fraction of strategic agents, leading to substantial variance in all metrics and a relatively high incidence of paper destruction. As $\varepsilon$ decays and more agents adopt the partially trained policy (episodes 100--300), the DRL network begins to identify and exploit patterns in the state space that correlate with long-run utility gains—specifically, recognizing when the immediate positional benefit of an ultimatum is outweighed by the expected reputational penalty and the loss of future collaboration opportunities. By episode 300, the policy has largely converged, and subsequent episodes (300--500) serve primarily to refine the Q-value estimates and to stabilize performance under the high-strategic-density regime.

It is worth noting that the training procedure did not employ any form of opponent modeling or multi-agent coordination beyond the shared replay buffer and the gradual conversion of greedy to strategic agents. Each strategic agent learned independently from its own trajectory data, with the DRL integrator treating the other agents (both greedy and strategic) as part of the stochastic environment. This design choice reflects the practical constraint that, in real academic settings, researchers typically do not have access to detailed models of their collaborators' decision processes. The fact that the policy nonetheless converges to a high-performance, low-destruction regime underscores the robustness of the DRL approach: by directly optimizing cumulative discounted utility in a sufficiently rich state representation, agents implicitly learn to cooperate and to avoid triggering the destructive equilibria that plague myopic strategies.

\begin{table}[H]
\centering
\caption{Utility distribution statistics for strategic and greedy agents across 
all population compositions. $\mu_{\text{Strat}}$ ($\sigma_{\text{Strat}}$) and 
$\mu_{\text{Greedy}}$ ($\sigma_{\text{Greedy}}$) denote the mean and standard 
deviation of cumulative utilities for each subpopulation. \textit{Advantage}: 
absolute utility premium of strategic over greedy agents 
($\mu_{\text{Strat}} - \mu_{\text{Greedy}}$). \textit{Rel.\,\%}: corresponding 
relative premium as a percentage of the greedy mean. Dashes indicate subpopulations 
absent from a given configuration (0\% and 100\% strategic).}
\label{tab:mixed_results}
\begin{tabular}{ccccccc}
\hline \hline
\textbf{Strategic \%} & \textbf{$\mu_{\text{Strat}}$} & \textbf{$\sigma_{\text{Strat}}$} & \textbf{$\mu_{\text{Greedy}}$} & \textbf{$\sigma_{\text{Greedy}}$} & \textbf{Advantage} & \textbf{Rel. \%} \\
\hline \hline
0   & ---      & ---      & 27{,}396 & 10{,}841 & ---    & ---  \\
10  & 35{,}010 & 13{,}215 & 26{,}765 & 10{,}875 & 8{,}246  & 30.8 \\
20  & 33{,}455 & 12{,}280 & 26{,}457 & 10{,}840 & 6{,}998  & 26.5 \\
30  & 33{,}204 & 12{,}565 & 26{,}296 & 10{,}797 & 6{,}908  & 26.3 \\
40  & 32{,}241 & 12{,}035 & 26{,}123 & 10{,}857 & 6{,}118  & 23.4 \\
50  & 32{,}142 & 12{,}320 & 25{,}786 & 10{,}938 & 6{,}356  & 24.6 \\
60  & 31{,}691 & 11{,}946 & 25{,}663 & 10{,}638 & 6{,}028  & 23.5 \\
70  & 31{,}499 & 12{,}131 & 25{,}765 & 10{,}641 & 5{,}734  & 22.3 \\
80  & 31{,}211 & 11{,}714 & 25{,}858 & 10{,}749 & 5{,}353  & 20.7 \\
90  & 30{,}974 & 11{,}917 & 25{,}370 & 10{,}754 & 5{,}604  & 22.1 \\
100 & 30{,}759 & 11{,}875 & ---      & ---      & ---    & ---  \\
\hline \hline
\end{tabular}
\end{table}

\begin{table}[H]
\centering
\caption{Ultimatum dynamics by agent type and population composition. \textit{Init.\ rate}: mean ultimatums raised per paper participated, per agent. \textit{Pr(Acc/Wd/Term)}: probability of acceptance, withdrawal, or termination conditional on initiator type. \textit{Resp.\ accept}: fraction of ``yes'' votes when an agent of this type is a responder. \textit{Restraint}: among failed ultimatums (not accepted), the fraction that ended in voluntary withdrawal rather than paper destruction. \textit{Destr.\ rate}: fraction of all ultimatums by this type that resulted in paper destruction.}
\label{tab:ultimatum_dynamics}
\resizebox{\textwidth}{!}{%
\begin{tabular}{c|cc|ccc|ccc|cc|cc|cc}
\hline
 & \multicolumn{2}{c|}{\textbf{Init.\ Rate}} 
 & \multicolumn{3}{c|}{\textbf{Outcome (Strat.\ Init.)}} 
 & \multicolumn{3}{c|}{\textbf{Outcome (Greedy Init.)}} 
 & \multicolumn{2}{c|}{\textbf{Resp.\ Accept}} 
 & \multicolumn{2}{c|}{\textbf{Restraint}} 
 & \multicolumn{2}{c}{\textbf{Destr.\ Rate}} \\
\textbf{Strat.\%} 
 & S & G 
 & Acc & Wd & Term 
 & Acc & Wd & Term 
 & S & G 
 & S & G 
 & S & G \\
\hline
10 & 0.292 & 0.296 & 0.39 & 0.61 & 0.00 & 0.42 & 0.46 & 0.13 & 0.51 & 0.73 & 1.00 & 0.78 & 0.00 & 0.13 \\
20 & 0.299 & 0.296 & 0.36 & 0.64 & 0.00 & 0.39 & 0.48 & 0.13 & 0.54 & 0.72 & 1.00 & 0.79 & 0.00 & 0.13 \\
30 & 0.306 & 0.297 & 0.34 & 0.66 & 0.00 & 0.38 & 0.49 & 0.13 & 0.55 & 0.71 & 1.00 & 0.79 & 0.00 & 0.13 \\
40 & 0.309 & 0.300 & 0.32 & 0.68 & 0.00 & 0.37 & 0.50 & 0.14 & 0.56 & 0.70 & 1.00 & 0.79 & 0.00 & 0.14 \\
50 & 0.309 & 0.303 & 0.32 & 0.68 & 0.00 & 0.36 & 0.50 & 0.14 & 0.57 & 0.69 & 1.00 & 0.79 & 0.00 & 0.14 \\
60 & 0.310 & 0.301 & 0.32 & 0.68 & 0.00 & 0.37 & 0.49 & 0.14 & 0.59 & 0.68 & 1.00 & 0.78 & 0.00 & 0.14 \\
70 & 0.312 & 0.302 & 0.32 & 0.68 & 0.00 & 0.36 & 0.50 & 0.14 & 0.59 & 0.67 & 1.00 & 0.79 & 0.00 & 0.14 \\
80 & 0.311 & 0.306 & 0.32 & 0.68 & 0.00 & 0.37 & 0.50 & 0.13 & 0.60 & 0.67 & 1.00 & 0.80 & 0.00 & 0.13 \\
90 & 0.315 & 0.308 & 0.32 & 0.68 & 0.00 & 0.37 & 0.50 & 0.13 & 0.60 & 0.67 & 1.00 & 0.80 & 0.00 & 0.13 \\
100 & 0.314 & 0.000 & 0.32 & 0.68 & 0.00 & 0.00 & 0.00 & 0.00 & 0.60 & 0.00 & 1.00 & 0.00 & 0.00 & 0.00 \\
\hline
\end{tabular}%
}
\end{table}

\begin{table}[ht!]
\centering
\caption{Productivity distribution statistics (Lotka-style analysis). For each population composition, we report the mean number of completed papers per active author, the power-law tail exponent $\hat{\alpha}$ and its $R^2$, the lognormal fit parameters $\hat{\mu}$ and $\hat{\sigma}$ with $R^2$, and the share of total production attributable to the top 10\% most productive authors. Power-law exponents are estimated via OLS on the log--log CCDF; lognormal parameters via MLE.}
\label{tab:lotka_stats}
\resizebox{\textwidth}{!}{%
\begin{tabular}{c|cccc|cccc|cccc}
\hline
 & \multicolumn{4}{c|}{\textbf{All Agents}} & \multicolumn{4}{c|}{\textbf{Strategic}} & \multicolumn{4}{c}{\textbf{Greedy}} \\
\textbf{Strat.\%} & $\bar{x}$ & $\hat{\alpha}$ (PL) & $\hat{\mu},\hat{\sigma}$ (LN) & Top10\% & $\bar{x}$ & $\hat{\alpha}$ & $\hat{\mu},\hat{\sigma}$ & Top10\% & $\bar{x}$ & $\hat{\alpha}$ & $\hat{\mu},\hat{\sigma}$ & Top10\% \\
\hline
10 & 47.3 & 9.32 (0.98) & 3.79,0.43 (0.77) & 16.3\% & 62.0 & 9.17 (0.95) & 4.09,0.29 (0.97) & 15.3\% & 45.7 & 9.67 (0.98) & 3.76,0.43 (0.73) & 15.9\% \\
20 & 47.9 & 9.51 (0.99) & 3.80,0.43 (0.76) & 16.2\% & 59.5 & 10.01 (0.98) & 4.05,0.27 (1.00) & 15.2\% & 45.0 & 9.14 (1.00) & 3.74,0.44 (0.71) & 15.7\% \\
30 & 49.0 & 9.64 (0.98) & 3.83,0.39 (0.84) & 16.1\% & 59.5 & 10.21 (0.99) & 4.06,0.24 (0.99) & 15.0\% & 44.4 & 9.50 (0.99) & 3.74,0.40 (0.76) & 15.3\% \\
40 & 49.6 & 10.32 (0.98) & 3.85,0.37 (0.82) & 15.8\% & 58.1 & 11.12 (0.99) & 4.03,0.25 (1.00) & 14.7\% & 44.0 & 9.78 (0.98) & 3.73,0.39 (0.76) & 15.3\% \\
50 & 50.6 & 10.47 (0.98) & 3.87,0.36 (0.84) & 15.6\% & 57.5 & 11.48 (0.99) & 4.03,0.22 (0.98) & 14.7\% & 43.7 & 9.50 (0.99) & 3.72,0.40 (0.75) & 15.2\% \\
60 & 51.1 & 10.68 (0.95) & 3.89,0.33 (0.84) & 15.2\% & 56.5 & 12.52 (0.97) & 4.01,0.22 (0.99) & 14.5\% & 43.1 & 10.21 (0.98) & 3.71,0.39 (0.70) & 14.8\% \\
70 & 52.3 & 11.08 (0.97) & 3.92,0.28 (0.94) & 14.9\% & 56.2 & 12.39 (0.98) & 4.01,0.21 (0.98) & 14.3\% & 43.3 & 10.26 (0.99) & 3.73,0.33 (0.83) & 14.6\% \\
80 & 53.2 & 11.56 (0.98) & 3.95,0.25 (0.97) & 14.6\% & 55.6 & 12.81 (0.99) & 4.00,0.20 (0.97) & 14.2\% & 43.9 & 9.32 (0.99) & 3.74,0.33 (0.86) & 14.6\% \\
90 & 54.0 & 11.17 (0.99) & 3.96,0.23 (0.99) & 14.4\% & 55.2 & 11.88 (1.00) & 3.99,0.20 (0.90) & 14.2\% & 43.6 & 9.39 (0.98) & 3.73,0.32 (0.86) & 14.5\% \\
100 & 54.8 & 10.90 (0.98) & 3.98,0.20 (0.75) & 14.1\% & 54.8 & 10.90 (0.98) & 3.98,0.20 (0.75) & 14.1\% & --- & --- & --- & --- \\
\hline
\end{tabular}%
}
\end{table}

\end{document}